\documentclass[preprint,aps]{revtex4}
\usepackage{graphicx}
\usepackage{eepic}
\usepackage{epic}
\usepackage{axodraw}

\newcommand{\beq}{\begin{equation}}
\newcommand{\eeq}{\end{equation}}
\def\bea{\begin{eqnarray}}
\def\eea{\end{eqnarray}}
\def\nn{\nonumber}

\begin{document}

\preprint{}

\title{ Systematic Analysis of $B\rightarrow K \pi l^+ l^- $ Decay \\
        through Angular Decomposition}
\author{C.~S.~ Kim}
\email{cskim@yonsei.ac.kr}
\affiliation{Department of Physics and IPAP, Yonsei
              University, Seoul 120-749, Korea.}

\author{Tadashi~ Yoshikawa}
\email{tadashi@eken.phys.nagoya-u.ac.jp}
\affiliation{
            Department of Physics, Nagoya University,
            Nagoya 464-8602, Japan. }

\begin{abstract}
\noindent We investigate systematically how to extract new physics contributions in
$B\rightarrow K\pi l^+ l^- $ decay by using the angular
decomposition.  The decomposition will enable us to define not only several CP averaged
forward-backward (FB) asymmetries but also
the direct CP asymmetry and the time-dependent mixing induced CP asymmetry
for each FB asymmetry newly defined in the general 4 body angular space.
The decay process  involves several intermediate vector and scalar resonances as sources of
strong phase difference through interference,
therefore, one can expect largely enhanced CP asymmetries, if there exists any new physics with
weak CP phases.
The combined analysis of the
FB and CP asymmetries will give us fruitful information about new physics contributions in detail.
\\

\end{abstract}
\maketitle

\section{Introduction}

One of the most important aims of present $B$ factories and future super-$B$ factories \cite{superB}
would be  to find out
evidences of new physics beyond the standard model (SM).
Indeed, to search for new physics evidence, we have investigated many penguin dominant processes,
which have loop diagrams as a leading contribution. A couple of years ago, we had two such definite evidences:
excitingly large discrepancies in CP asymmetries for $b\rightarrow s\bar{q}q $
decays, eg. $B\rightarrow \phi K$ \cite{phiK,LS}, and
smaller but much  unexpected  discrepancies in $B\rightarrow K\pi $ decays,
so called ``$B\rightarrow K\pi$ puzzle'' \cite{lipkin,yoshi,BFRS,GR,C-L,recent-PP-KP,CPEWP}
between theoretical predictions within the SM and the
experimental data.

For ``$B\rightarrow K\pi$ puzzle'', the experimental data had shown three large discrepancies from
the SM predictions of the branching ratios and CP asymmetries.
One of them is the difference between the ratios of
branching ratio for charged $B$ decays ($R_c$) and for neutral $B$ decays
($R_n$). The second one is that between the direct CP asymmetries for $B^{+}
\rightarrow K^{+}\pi^0 $ and  $B^0
\rightarrow K^{+}\pi^- $. The third one is that between the weak phase $\sin2\phi_1$
extracted from the time-dependent CP asymmetry of $B^{0} \rightarrow K^{0}\pi^0 $
and of $B^{0} \rightarrow J/\psi K_s $.
Main contribution of all $B\rightarrow K\pi $
modes comes from $b \to s$ QCD penguin processes, therefore, the sub-leading electro-weak (EW)
penguin type new physics contribution  has been considered as the most
plausible source to explain those three discrepancies \cite{LS,lipkin,yoshi,BFRS,GR,C-L,recent-PP-KP,CPEWP}.
Recently, the $R_c - R_n$ discrepancy has been disappearing
but the other two differences  seem to be still remaining. It could mean that
sizable parameter space for new physics is still valid in these decay
modes.  One of such possibilities is  new physics with large CP phases in EW
penguins \cite{yoshi,CPEWP}.

Investigation of the CP phase in EW penguin processes is
very important to
check the SM and confirm the discrepancies in $B\rightarrow K\pi$
modes at the same time. To do so, the semi-leptonic rare decays $b\rightarrow s l^+ l^-$,
which are pure EW penguin processes with less hadronic
uncertainty than the hadronic $B$ decays, can be
the best modes to search for the evidence of new CP phase in
EW penguin diagram.
$B\rightarrow K^* l^+ l^- $ is a $b \to s$ EW
transition process, so that the penguin vertex does not have large weak
phase within the SM. Therefore, we have to either confirm the feature about only small CP phase
coming from the CKM Matrix \cite{CKM}, or  search for some evidences
of new physics with large CP phases beyond the SM.

Several semi-leptonic rare decays, $B\rightarrow M l^+ l^- $ modes,
have been measured \cite{EXbvll} and they will
provide very useful information of new physics in  EW
penguin contributions \cite{LMS,Ali1,THbvll,Ali1.5,Ali2,hurth,bvll}.
To analyze the source of new physics, we can try to
extract a few hints by using radiative decays $b\rightarrow s \gamma$.
However, due to the absence of large strong phase in the decay, the CP asymmetry of $b\rightarrow s \gamma$
would be very tiny. To investigate the new contributions to CP phase
in $b\rightarrow s \gamma $,
we seem to need new experimental technique.
(For example, using photon conversion technique \cite{Ntech} one can
determine the parameters with new CP phase.)

The rare decays
$b\rightarrow s l^+ l^- $\cite{bvll,CPFB,CPbvll} can be much more
interesting process because these decays are including possibly
large strong phases induced by the $(c\bar{c})$ intermediate resonance states.
Furthermore, for the decays of $B \to M [\to K \pi]~ l^+ l^-$~~($M=K^*, K_0^*(800),...$),
if we do not constrain the invariant mass of
$K$-$\pi$ system, there may exist several intermediate mesons, $M$, contributing to
$B \to K \pi l^+ l^-$ decays. Therefore, through the interference we may induce large strong phases,
which results possibly large CP violations if there is any new physics with
weak CP phases.

We are interested in CP asymmetries and forward-backward (FB) asymmetries\cite{CPFB} for $B \to K \pi  l^+ l^-$
decays to extract  information on possible new CP phases in $b \to s$ EW
penguin transitions.
Here the final state are including both
CP odd and CP even so that it may be  slightly difficult to consider
the CP asymmetries. If we consider the time-dependent CP
asymmetry, it cannot be even defined under this condition including both
CP odd and even states.
Hence we have to decompose the mode by using angular analysis.
{}From the decomposition, one can define many observables and  CP
asymmetries so that one may be able to obtain
fruitful information. Some of them are very sensitive to
strong or EW phases.
Some of them are from interference contributions between CP odd
and CP even modes so that the CP asymmetry may be enhanced.
Therefore, here we consider the angular analysis of 4 body decays
$B\rightarrow K\pi l^+ l^-$ \cite{bkpll,KKLM} and the CP asymmetries
through the angular  decomposition.

The important points in this work are:
\begin{itemize}
\item{Angular decomposition of the decay rate, forward-backward
      asymmetries \cite{Ali1} and CP asymmetries\cite{CPFB} are investigated.}
\item{Dependence of strong phases from several resonances in
       dilepton part and $K\pi$ part to CP asymmetries.
       If the intermediated states are including several
       meson states in addition to vector meson $K^*$, the interferences may have
       an important role as a source of strong phase difference, which
       is one of the conditions to enhance the CP asymmetry. }
\item{Using model-independent analysis \cite{FKMY,Aliev},
       the new physics information
       can be clearly classified. }
\end{itemize}

This paper is organized as follows. In section 2, we show several
definitions to calculate $B\rightarrow K\pi l^+ l^-$ decays and
derive the branching ratio and the angular decomposition
from the most general 4-fermi interaction. And we define the direct
and indirect CP asymmetries of each decomposed FB asymmetry. In
section 3, several figures of FB asymmetries and the CP asymmetries
are plotted under some conditions.
In section 4, the case with scalar resonance in addition to $K^*$ is discussed.
The interference effect may make a new source of strong phase
difference to enhance the CP asymmetries. Section 5 is devoted to our
summary.

\section{Theoretical details of $B\rightarrow K^* [\to K \pi]~ l^+ l^-$ decays}

\begin{figure}[b]
\begin{minipage}[c]{5.0in}
{\includegraphics[scale=1]{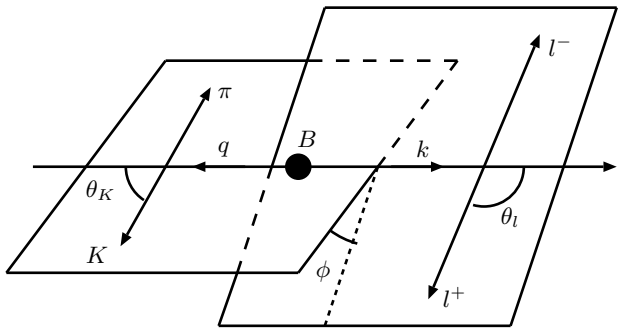}}
\end{minipage}
\caption{ Definition of the kinetic variables and the angles in
$B\rightarrow K\pi l^+l^- $ decay.  Here $q$ is the intermediate
meson momentum and $k$ is the intermediate photon momentum. }
\end{figure}

To describe systematically  the general 4 body decay,
$B(P_B)\rightarrow M [\to K(P_K)+\pi(P_{\pi})]+ l^+(P_+)+ l^-(P_-)$, where $P_x$
is the momentum of each particle,
we define two kinetic variables, $s$ and $z$, and three angles,
$\theta_l, \theta_K$ and $\phi $ \cite{bkpll,KKLM}. (See Fig.1.)
Here $q$ is the momentum of
intermediate state $M$, $i.e.$ the sum of the momenta of $K$ and
$\pi $ mesons, and  $s = q^2 = (P_K + P_\pi)^2$. And $z$ is
defined as the
square of invariant mass of dilepton, $z=k^2 =(P_+ + P_-)^2$.
$\theta_l$ is an angle  between the momentum direction of $l^+$ and that of the intermediate photon
(or opposite direction of the intermediate meson) at the center of mass (CM)
system of $l^+$ and $l^-$. $\theta_K$ is an
angle between $K$ direction and the intermediate meson ($M$) direction at CM
system of $K$ and $\pi $.
And $\phi $ is an angle between the two decay planes at $B$ rest
frame. Using the three angles, we can decompose the decay,
$B\rightarrow K\pi l^+ l^-$, completely.
For simplicity, we assume the leptons and $K$ and $\pi$ mesons all massless.

As a systematic analysis, we start from the most general 4-fermi
lagrangian \cite{FKMY}. It consists of
12 most general independent four-Fermi interactions,
\bea
\bar{\cal M}(b\rightarrow sl^+l^-)
    = \frac{G_F \alpha }{\sqrt{2}\pi } V_{ts}^*V_{tb} \left[ \right.&-&2
    \bar{s}i\sigma_{\mu\nu}\frac{k^\nu}{k^2}(C_{7}m_b P_R +
    C_{7}^\prime m_b P_L)b \bar{l}\gamma^\mu l \nn \\
    &+& \bar{s}\gamma_\mu(C_9 P_L + C_9^\prime P_R) b
    \bar{l}\gamma^\mu l \nn \\
  &+& \bar{s}\gamma_\mu (C_{10} P_L +
    C_{10}^\prime P_R ) b \bar{l}\gamma^\mu \gamma_5 l \nn \\
  &+& \bar{s}(C_{SS} + C_{AS}\gamma_5 )b \bar{l} l \nn \\
  &+& \bar{s}(C_{SA} + C_{AA}\gamma_5 )b \bar{l} \gamma_5 l \nn \\
  &+& \left. \bar{s}\sigma_{\mu\nu }b \bar{l}\left( C_T \sigma^{\mu\nu } + i
    C_{TE} \sigma_{\alpha \beta }~\epsilon^{\mu\nu\alpha\beta}
    \right) l ~~\right],
\eea
where $C_X$ is the coefficient for each four-Fermi
interaction. $C_7$, $C_9$ and $C_{10}$ correspond to the 3 parameters
in the SM. The other coefficients will show the contributions from the
interactions beyond the SM. $C_7^\prime $ within the SM is
suppressed by $m_s/m_b$ factor so that its contribution is estimated as tiny.
If  right-handed currents as new physics interactions exist,
$C_7^\prime, C_9^\prime, C_{10}^\prime $ will show non negligible contributions.
$C_{SS},C_{SA},C_{AS}$ and $C_{AA}$ come from scalar type new physics
interactions. And $C_{T}$ and $C_{TE}$ show the tensor type
contributions.
In general we can define a new CP phase as $e^{i\phi_x^{(\prime )}} $
for each Wilson coefficient $C_x^{(\prime )}$ in Eq. (1).

To calculate the process
$\{\bar{B}^0, B^- \}\rightarrow M~(e.g.~K^*) ~[\rightarrow K \pi]~ l^+ l^- $,
we are using the
following parametrization \cite{Ali2} for the matrix element of the hadronic part:
\bea
<{K}^*|\bar{s}\gamma_\mu P_{L,R} b |{B}> &=& i \epsilon_{\mu\nu\rho\sigma}
       \varepsilon^{\nu *}q^\rho k^{\sigma}\frac{V(z)}{m_B + m_{K^*}}
       \mp \left\{ \varepsilon_\mu^* (m_B + m_{K^*}) A_1(z) \right.
          \nn \\
   &-&\left.
         (\varepsilon^* \cdot k)(2 q + k)_\mu
       \frac{A_2(z)}{m_B+m_{K^*}}
        - \frac{2 m_{K^*}}{z}(\varepsilon^*\cdot k) k_\mu
       [A_3(z)-A_0(z)]\right\},\nn \\
                               \\[3mm]
<K^*|\bar{s}i\sigma_{\mu\nu}k^\nu P_{R,L} b |{B}>
   &=& -i \epsilon_{\mu\nu\rho\sigma}
       \varepsilon^{\nu *}q^\rho k^{\sigma} T_1(z)
       \pm \left\{ [\varepsilon_\mu^* (m_B + m_{K^*})
                 -(\varepsilon\cdot k)(2 q + k)_\mu] T_2(z) \right.
          \nn \\
   & &\left.-
         (\varepsilon^* \cdot k)[k_\mu - \frac{z}{m_B^2-m_{K^*}^2}(2 q + k)_\mu
       T_3(z) ]\right\}, \\[3mm]
<K^*|\bar{s}b |B>&=&0, \\[3mm]
<K^*|\bar{s}\gamma_5 b |{B}>&=&-\frac{2m_{K^*}}{m_B}[ \varepsilon^*
       \cdot k A_0(z) ],\\[3mm]
<K^*|\bar{s}\sigma_{\mu\nu} b |{B}> &=& \epsilon_{\mu\nu\rho\sigma}
       [ -\varepsilon^{\lambda *}(2q +k)^\sigma T_1(z)
         + \frac{m_B^2 - m_{K^*}^2}{z}\varepsilon^{\lambda *} k^\sigma
       \{T_1(z)-T_2(z)\} \nn \\
   & & -\frac{2}{z} \varepsilon^*\cdot k q^\lambda k^\sigma
           \{ T_1 - T_2 - \frac{z}{m_B^2 - m_{K^*}^2} T_3 \}],
 \eea
where $\varepsilon^\nu $ is the polarization vector of $K^*$
       meson, and $P_{L,R}$ is the projection operator  $P_{L,R} = (1
       \mp \gamma_5)/2 $.
For form factors $V(z),A_i(z), T_i(z)$, we follow
       the definition of Ref. \cite{Ali2}.
Considering the current conservation of the leptonic part, the terms
including $k_\mu $ will disappear.
For the decay process of the intermediate vector meson, $K^*
       \rightarrow K \pi $, the contribution  is replaced as follows \cite{GodinaNava:1995jb};
\bea
<K\pi|K^*><K^*| = g_{K\pi} (P_K - P_\pi )_\alpha
          \frac{g^{\alpha\nu}- \frac{q^\alpha q^\nu}{q^2}}{G} = g_{K\pi} (P_K - P_\pi )_\alpha
          \frac{g^{\alpha\nu}- \frac{q^\alpha q^\nu}{q^2}}{m_{K^*}^2 -
          q^2 - i m_{K^*} \Gamma_{K^*}}
\eea
where $g_{K\pi} $ is the decay constant and $\Gamma_{K^*} $ is the decay
width of ${K^*}$ meson.

Under our parametrization,
the branching ratio can be expressed as
\bea
{\cal B}({B}\rightarrow K\pi ll) &= & \frac{  \int ds dz Y
   B(s,z)}{\Gamma_B},   \nn
\eea
where
\bea
B(s,z) &\equiv& \int d\phi d(\cos\theta_K) d(\cos\theta_l)
(\Gamma_1 + \Gamma_2 + \Gamma_3 +\Gamma_4 + \Gamma_5 + \Gamma_6 +
   \Gamma_7) \nn \\
       &=& \int d\phi d(\cos\theta_K) d(\cos\theta_l)\Gamma_1~, \\[3mm]
Y &=& \frac{G_F^2 \alpha^2 |V_{tb}^*V_{ts}|^2 L[s,z]}{128\times 512 \pi^8}.
\eea
$\Gamma_B$ is  $B$-meson total decay width, and
\bea
\Gamma_1 \equiv \frac{g_{K\pi}^2}{|G|^2}
      & &\left\{ \left( |C_9^{eff}(z) - C_9^\prime|^2
                      + |C_{10} - C_{10}^\prime|^2 \right) \right. \nn
                      \\
      & & ~~~~~~~~~~~~~~~  \times
          \frac{1}{8 ( m_B + m_{K^*} )^2 }
          \left.\left[ (m_B+ m_{K^*})^4 (4 s z B_1 + L_0^2 B_3 )~|A_1|^2
                                              \right. \right. \nn \\
      & & ~~~~~~~~~~~~~~~~~~~~~~~~~~~~~~~~~~~~~~~~~~~~
            + L^4 B_3 ~|A_2|^2 \nn \\
      & &~~~~~~~~~~~~~~~~~~~~~~~~~~~~~~~~~~~~~~~~~~~~
         - \left.  2 L^2 L_0 ( m_B + m_{K^*})^2 B_3 ~(A_1 A_2)  \right] \nn \\
      & & + 4 |C_7 - C_7^\prime |^2 \frac{m_b^2}{8
      z^2 } \left[ \left( 4 (m_B^2 - m_{K^*}^2 )^2 s z B_1
          \right.\right. \nn \\
      & & ~~~~~~~~~~~~~~~~~~~~~~~
          \left. \left.
       +
         \left(L^2 - L_0 (m_B^2 - m_{K^*}^2 ) \right)^2 B_3 \right) ~|T_2|^2 \right. \nn \\
      & & ~~~~~~~~~~~~~~~~~~~~~~~
          + \frac{z^2  L^4}{(m_B^2 - m_{K^*}^2 )^2} B_3 ~|T_3|^2 \nn \\
      & & ~~~~~~~~~~~~~~~~~~~~~~~
          \left. + 2 \frac{z L^2}{(m_B^2 - m_{K^*}^2 )}
           (L^2 - L_0(m_B^2 - m_{K^*}^2)) B_3
      ~(T_2 T_3 ) \right]\nn \\
      & & + 4 Re\left((C^{eff}_9(z)^* - C_9^{\prime *})(C_7 -
                      C_7^\prime )\right) \nn \\
      & & ~~~~~~~~~~~~~~~  \times
           \frac{m_b}{16(m_B+m_{K^*}) z } \left[ (m_B+m_{K^*})^2 \left\{
            8 s z ( m_B^2 - m_{K^*}^2 ) B_1 \right. \right. \nn \\
      & & ~~~~~~~~~~~~~~~~~~~~~~~~~~~~~~~~~~~~~~~~~~~
          \left.       - 2 L_0 (L^2 - (m_B^2 - m_{K^*}^2) L_0 ) B_3 \right\}
      ~(A_1 T_2)
          \nn \\
      & & ~~~~~~~~~~~~~~~~~~~~~~~~~~~~~~~~~~~~~~~~~~~
          - 2 z L^2 L_0 B_3 ~(A_1 T_3)\nn \\
      & & ~~~~~~~~~~~~~~~~~~~~~~~~~~~~~~~~~~~~~~~~~~~
          + 2 L^2 (L^2 - (m_B^2 - m_{K^*}^2) L_0 ) B_3 ~(A_2 T_2) \nn \\
      & & ~~~~~~~~~~~~~~~~~~~~~~~~~~~~~~~~~~~~~~~~~~~
          \left.\left. + \frac{2 z L^4}{(m_B^2 - m_{K^*}^2)} B_3 ~(A_2 T_3)
          \right]
                \right.\nn \\
      & & +\left. \left( |C_9^{eff}(z)+C_9^\prime|^2 + |C_{10}+C_{10}^\prime|^2 \right)
              \frac{1}{2 ( m_B + ms )^2 } L^2 s z B_2 ~|V|^2 \right.\nn \\
      & & + 4 |C_7 + C_7^\prime |^2 \frac{m_b^2 s}{2 z } L^2 B_2 ~|T_1|^2
          \nn \\
      & & \left. \left.+4 Re\left((C^{eff}_9(z)^* +C_9^{\prime *}) (C_7 + C_7^\prime )\right)
            \frac{s m_b}{2(m_B+m_{K^*})} L^2 B_2 ~(T_1 V)
          \right.  \right. \\
      & &  \nn  \\
      & &+\left(|C_{AS}|^2 +|C_{AA}|^2 \right) \frac{m_{K^*}^2}{m_B^2} 2 z L^2
                   \cos^2\theta_K |A_0|^2 \nn \\
      & &+\left(|C_{T}|^2 +4|C_{TE}|^2 \right)\frac{8}{z}
          \left[ \{(m_B^2 - m_{K^*}^2 -
          z -L_0 )^2 ( L_0^2 S_1
         + 4 s z S_2) \right. \nn
          + 4 s z L^2 S_3 \}|T_1|^2  \nn \\
     & &  ~~~~~~~~~~~~~~~~~~~~~~~~~~~+\{(L^2 - L_0(m_B^2 - m_{K^*}^2))^2 S_1 + 4 s z (m_B^2 - m_{K^*}^2)^2
                      S_2 \} |T_2|^2 \nn \\
     & & ~~~~~~~~~~~~~~~~~~~~~~~~~~~~+\frac{z^2 L^4}{(m_B^2 - m_{K^*}^2
                      )^2} S_1 |T_3|^2 \nn \\
     & & ~~~~~~~~~~~~~~~~~~~~~~~~~~~~- \{(L_0 - m_B^2 + m_{K^*}^2 +z
                      )(L_0(L^2 - L_0(m_B^2 - m_{K^*}^2))
                      S_1 \nn \\
     & & ~~~~~~~~~~~~~~~~~~~~~~~~~~~~~~~~~~~~~~~~~
         - 4sz (m_B^2 - m_{K^*}^2)^2 S_2 \} 2 T_1 T_2 \nn \\
     & & ~~~~~~~~~~~~~~~~~~~~~~~~~~~~- \frac{z L^2 L_0}{m_B^2 -m_{K^*}^2 } \{ (L_0 - m_B + m_{K^*} +z )
                      S_1\} 2 T_1 T_3 \nn \\
     & & ~~~~~~~~~~~~~~~~~~~~~~~~~~~~+  \frac{z L^2 }{m_B^2 -m_{K^*}^2 } \{(L^2 - L_0(m_B^2 -
                      m_{K^*}^2))S_1 2 T_2 T_3 ], \nn
\eea
\bea
\Gamma_2 &\equiv& F_2(s,z) \sin^2\theta_K
          \cos\theta_l \\
       &=& \frac{g_{K\pi}^2}{|G|^2}  \sin^2\theta_K
          \cos\theta_l
                \nn \\
      &\times &\left\{ 2 Re\left( C^{eff}_9(z)^* C_{10}
                                 - C_9^{\prime *} C_{10}^\prime \right)
            L s z  ~(A_1 V)\right. \nn \\
      & & + 2 Re\left((C_{10}^* +C_{10}^\prime ) (C_7 - C_7^\prime )\right)
            m_b ( m_B - m_{K^*} ) s L  ~(V T_2) \nn \\
      & & \left. + 2 Re\left((C_{10}^* -C_{10}^\prime )(C_7 + C_7^\prime )\right)
            m_b ( m_B + m_{K^*} ) s L  ~(A_1 T_1) \right\},
\label{Gam2}
\\[5mm]
\Gamma_3  &\equiv& F_3(s,z) \cos\phi \sin2\theta_K
       \sin2\theta_l \\
     &=&\frac{g_{K\pi}^2}{|G|^2}
            \cos\phi \sin2\theta_K \sin2\theta_l \nn \\
      &\times& \left\{ \left( |C_9^{eff}(z) - C_9^\prime |^2 + |C_{10}
     - C_{10}^\prime |^2 \right) \right.
          \frac{\sqrt{s z }}{8}
          \left.\left[ (m_B+ m_{K^*})^2 L_0 ~|A_1|^2
                         \right. -  L^2  ~(A_1 A_2)  \right] \nn \\
      & & - 4 |C_7 - C_7^\prime |^2
                \frac{m_b^2 \sqrt{s z }}{8 z^2 }
           \left[ (m_B^2 - m_{K^*}^2 ) \left( L^2 - L_0
      (m_B^2 - m_{K^*}^2 ) \right)  ~|T_2|^2 + 2 z L^2
      ~(T_2 T_3 ) \right]\nn \\
      & & - 4 Re\left((C^{eff}_9(z)^* -C_9^{\prime *})(C_7 -
     C_7^\prime )\right) \nn \\
      & & ~~~~~~~~~~~~~~~~~~~~~~ \times
           \frac{m_b \sqrt{s z}}{16(m_B-m_{K^*}) z } \left[ (m_B^2-m_{K^*}^2)
           (L^2 - 2 (m_B^2 - m_{K^*}^2) L_0 )
      ~(A_1 T_2) \right.
          \nn \\
      & & ~~~~~~~~~~~~~~~~~~~~~~~~~~~~~~~~~~~~~~
          + z L^2  ~(A_1 T_3)
          \left. \left.+ L^2 (m_B^2 - m_{K^*}^2)  ~(A_2 T_2)
          \right] \right. \\
      & & + 8 \left(|C_T|^2 + 4 |C_{TE}|^2 \right)\frac{\sqrt{s z}}{z}
             \{ -L_0(L_0 - m_B^2 + m_{K^*}^2 +z )^2 |T_1|^2 \nn \\
      & & ~~~~~~~~~~~~~~~~~~~~~~~
           - 4 s z (m_B^2 - m_{K^*}^2 ) |T_2|^2 \nn \\
     & & ~~~~~~~~~~~~~~~~~~~~~~~ + (L^2 - 2 L_0 (m_B^2 - m_{K^*}^2 )) T_1
     T_2 \nn \\
     & & ~~~~~~~~~~~~~~~~~~~~~~~ + \frac{z L^2}{m_B^2 - m_{K^*}^2 }(L_0 -
     m_B^2 + m_{K^*}^2 +z ) T_1 T_3 \nn \\
     & & ~~~~~~~~~~~~~~~~~~~~~~~ + {z L^2} T_2 T_3
     \} \},
          \nn \\
\Gamma_4 &\equiv& F_4(s,z) \sin2\phi \sin^2\theta_K
     \sin^2\theta_l \\
     &=& \frac{g_{K\pi}^2}{|G|^2}
          \sin2\phi \sin^2\theta_K \sin^2\theta_l \nn \\
      &\times & \left\{ - Im\left( (C_9^{eff}(z)^* - C_9^{\prime *})(C_7 + C_7^\prime
      )\right) m_b s L (m_B + m_{K^*}) ~(A_1 T_1) \right. \nn \\
      & &  + Im\left( (C_9^{eff}(z)^* + C_9^{\prime *})(C_7 - C_7^\prime
      )\right) m_b s L (m_B - m_{K^*}) ~(V T_2) \nn \\
      & & \left. - Im(C_7^* C_7^\prime ) \frac{ 8 m_b^2 s L (m_b^2 -
      m_{K^*}^2)}{z} ~(T_1 T_2) \right\}, \\[5mm]
\Gamma_5 &\equiv& F_5(s,z)  \sin\phi\sin2\theta_K
     \sin2\theta_l \\
     &=& \frac{g_{K\pi}^2}{|G|^2}
                 \sin\phi\sin2\theta_K \sin2\theta_l \nn \\
      &\times & \left\{ Im\left( (C_9^{eff}(z)^* - C_9^{\prime *})(C_7 + C_7^\prime
      )\right) \frac{ m_b \sqrt{s z } L}{4(m_B + m_{K^*}) z} \left[
        L_0 (m_b^2 - m_{K^*}^2 ) ~(A_1 T_1) \right.
          - L^2 ~(A_2 T_1 ) \right] \nn \\
      & &  + Im\left( (C_9^{eff}(z)^* + C_9^{\prime *})(C_7 - C_7^\prime
      )\right) \frac{m_b \sqrt{s z } L}{4 (m_B - m_{K^*}) z}\left[
         \left(L^2 - (m_B^2-m_{K^*}^2 )L_0 \right) ~(V T_2) \right. \nn \\
      & & ~~~~~~~~~~~~~~~~~~~~~~~~~~~~~~~~~~~~~~~~ \left. -\frac{z L^2}{(m_B^2 - m_{K^*}^2)} ~(V T_3) \right]
       \\
      & & \left. - Im(C_7^* C_7^\prime ) \frac{ m_b^2 \sqrt{s z} L}{(m_b^2 -
      m_{K^*}^2) z^2 }\left[ (m_B^2 - m_{K^*}^2) \left( L^2 - (m_B^2 - m_{K^*}^2)
      L_0 \right) ~(T_1 T_2) \right.
         \left. + z L^2 ~(T1 T_3 ) \right] \right\}, \nn
\eea
\bea
\Gamma_6 &\equiv& F_6(s,z) \cos\phi\sin2\theta_K \sin\theta_l
     \\
       &=& \frac{g_{K\pi}^2}{|G|^2}
                 \cos\phi\sin2\theta_K \sin\theta_l \nn \\
      &\times &\left\{Re\left(C^{eff}_9(z)^*C_{10} - C_9^{\prime *}
       C_{10}^\prime \right)\frac{\sqrt{s
      z } L }{2(m_B+m_{K^*})^2}\left[ (m_B + m_{K^*})^2 L_0 ~(A_1 V) -
      L^2~(A_2 V ) \right] \right.\nn \\
     & & - Re\left((C_{10}^* + C_{10}^{\prime *})(C_7-C_7^\prime)\right) \frac{m_b L \sqrt{s
      z}}{2 (m_B+m_{K^*}) z} \left[ \left(L^2 - (m_B^2 - m_{K^*}^2)L_0\right)
      ~(V T_2) \right. \nn \\
     & & ~~~~~~~~~~~~~~~~~~~~~~~~~~~~~~~~~~~~~~~~~~~~~~~~~~~~~~
       + \left. \frac{L^3 m_b}{(m_B+m_{K^*})(m_B^2-m_{K^*}^2)}~(V T_3)
      \right]  \\
     & & \left. + Re\left((C_{10}^* - C_{10}^{\prime *})(C_7+C_7^\prime)\right) \frac{m_b L
      \sqrt{s z}}{2(m_B+m_{K^*}) z}\left[ (m_B+m_{K^*})^2 L_0 ~(A_1 T_1) - L^2
      ~(A_2 T_1) \right] \right\}, \nn
\\[5mm]
 \Gamma_7 &\equiv& F_7(s,z) \sin\phi\sin2\theta_K
       \sin\theta_l \\
     &=& \frac{g_{K\pi}^2}{|G|^2}
                 \sin\phi\sin2\theta_K \sin\theta_l \nn \\
      &\times &\left\{ Im\left((C_{10}^* - C_{10}^{\prime *})(C_7-C_7^\prime )\right)
        \frac{m_b L^2 \sqrt{s z } }{2 (m_B-m_{K^*}) z} \left[
        (m_B^2 - m_{K^*}^2)~(A_1 T_2)
        + z ~(A_1 T_3) \right. \right.\nn \\
     & & ~~~~~~~~~~~ \left.  - (m_B - m_{K^*})^2 ~(A_2 T_2)\right] \nn \\
     & & \left. + 8 Re\left((C_{S}^* C_T - 2 C_A^* C_{TE}\right)
           \frac{ m_{K^*}\sqrt{s z} L^2}{m_B} T_1 A_0 \right\},
\label{Gam7}
\eea
where
\bea
L &\equiv& \sqrt{ (s - z)^2 - 2 m_B^2 (s + z ) + m_B^4 }, \\
L_0 & \equiv& \sqrt{L^2 + 4 s z } = m_B^2 - s - z .
\eea
Angular functions in $\Gamma_1$ are
\bea
B_1 &=& \sin^2\theta_K - \cos^2\phi \sin^2\theta_K \sin^2\theta_l , \\
B_2 &=& \sin^2\theta_K - \sin^2\phi \sin^2\theta_K \sin^2\theta_l , \\
B_3 &=& \cos^2\theta_K\sin^2\theta_l , \\
S_1 &=& \cos^2\theta_K \cos^2\theta_l ,\\
S_2 &=& \sin^2\theta_K \sin^2\theta_l \cos^2\phi ,\\
S_3 &=& \sin^2\theta_K \sin^2\theta_l \sin^2\phi .
\eea

After integrating out whole angular space, we get the values of
$\Gamma_2$,..,$\Gamma_7$  becoming zero,
due to  the canceling angular dependence with an over-all factor of $F_{2,..,7}(s,z)$.
However, partial angular integration asymmetries becoming non-zero values, like FB asymmetries,
can give us possibly very important information on new physics contributions.
For each $\Gamma_i$, we can define new observable of FB asymmetry
defined in Eqs. (\ref{FB2})--(\ref{FB7}).
$E.g.$, $\Gamma_2$ is proportional to $\sin^2\theta_K \cos\theta_l $, so that the term appears as
FB asymmetry of leptons.
$\Gamma_4$ becomes an asymmetry for the angle $\phi$ between $0$ to $\frac{\pi}{2}$ and
$\frac{\pi}{2}$ to $\pi$. $\Gamma_6$ and $\Gamma_7$ show the double
asymmetries for $K$ meson and left-right or up-down asymmetry for the angle
$\phi$.  $\Gamma_3$ and $\Gamma_5$ show the triple FB
asymmetries  for $K$ and leptons.

Now we define the angular integration operators $FB_i$ which operate to
$\Gamma_{total} = \sum_{i=1}^7 \Gamma_i $, in order to extract the FB
asymmetries $FB_i~ \Gamma_i ~(i=2-7)$, as follow:
\bea
FB_{2}~\Gamma_{total}&\equiv&\int_0^{2\pi }d\phi \int_0^{\pi}\sin\theta_K d\theta_K
           \left(\int^{\frac{\pi}{2}}_{0}
               - \int^{\pi}_{\frac{\pi}{2}}\right) \sin\theta_l d\theta_l
           \Gamma_{2} =  \frac{8\pi}{3} F_2(s,z),
\label{FB2}
\\
FB_{3}~\Gamma_{total}&\equiv&
             \left(\int_{-\frac{\pi}{2}}^{\frac{\pi}{2}}
                    - \int^{\frac{3\pi}{2}}_{\frac{\pi}{2}}\right)d\phi
             \left(\int_{0}^{\frac{\pi}{2}}
                    - \int_{\frac{\pi}{2}}^\pi \right)
                            \sin\theta_K d\theta_K
             \left(\int^{\frac{\pi}{2}}_{0}
                    - \int^{\pi}_{\frac{\pi}{2}}\right)
                            \sin\theta_l d\theta_l
              \Gamma_{3} =  \frac{64}{9} F_3(s,z), \nn \\
                                                          \\
FB_{4}~\Gamma_{total} &\equiv& \left(\int_{0}^{\frac{\pi}{2}}
                    - \int^{{\pi}}_{\frac{\pi}{2}}\right)d\phi
               \int_0^\pi
                 \sin\theta_K d\theta_K \sin\theta_l d\theta_l
              \Gamma_{4} =  \frac{32}{9} F_4(s,z), \\
FB_{5}~\Gamma_{total} &\equiv& \left(\int_{0}^{{\pi}}
                    - \int^{2\pi}_{\pi}\right)d\phi
             \left(\int_{0}^{\frac{\pi}{2}}
                    - \int_{\frac{\pi}{2}}^\pi \right)
                            \sin\theta_K d\theta_K
             \left(\int^{\frac{\pi}{2}}_{0}
                    - \int^{\pi}_{\frac{\pi}{2}}\right)
                            \sin\theta_l d\theta_l
              \Gamma_{5} =  \frac{64}{9} F_5(s,z), \nn \\
                                                         \\
FB_{6}~\Gamma_{total}&\equiv&
             \left(\int_{-\frac{\pi}{2}}^{\frac{\pi}{2}}
                    - \int^{\frac{3\pi}{2}}_{\frac{\pi}{2}}\right)d\phi
             \left(\int_{0}^{\frac{\pi}{2}}
                    - \int_{\frac{\pi}{2}}^\pi \right)
                            \sin\theta_K d\theta_K
              \int^{\pi}_{0}
                            \sin\theta_l d\theta_l
              \Gamma_{6} =  \frac{16\pi}{9} F_6(s,z), \\
FB_{7}~\Gamma_{total}&\equiv&
             \left(\int_{0}^{{\pi}}
                    - \int^{\pi}_{2\pi}\right)d\phi
             \left(\int_{0}^{\frac{\pi}{2}}
                    - \int_{\frac{\pi}{2}}^\pi \right)
                            \sin\theta_K d\theta_K
              \int^{\pi}_{0}
                            \sin\theta_l d\theta_l
              \Gamma_{7} =  \frac{16\pi}{9} F_7(s,z).
\label{FB7}
\eea
Because $K$--$\pi$ system of $B\rightarrow M (\to K \pi) l^+l^-$ decay
is a mixture of CP-even and CP-odd modes,
investigating CP asymmetry of the decay is not so simple, even though
it is very important to extract
new physics information beyond the SM.
However, by combining  CP asymmetries with previously defined FB
asymmetries, we can clearly separate
the final states with CP eigen-mode.
The mixing induced time-dependent CP asymmetry can be also considered similarly.
The CP eigen-mode for each $\Gamma_i$ is follows:
$\Gamma_2, \Gamma_4, \Gamma_5$ and $\Gamma_6$ are CP odd, and
$\Gamma_3, \Gamma_7$ are CP even.

The CP averaged FB asymmetries can be defined as follows:
\bea
A^{FB_i}(s,z) = \frac{FB_i[ \eta_{cp} \bar{\Gamma}_i + \Gamma_i
]}{\bar{B}(s,z) + B(s,z)},
\label{AFB}
\eea
where $\eta_{CP} = +1$ for CP even, and $-1$ for CP odd.
$\bar{B}(s,z)$ and $\bar{\Gamma_i}$ represent for the CP conjugate $B$
meson decays.
Usual definition of FB asymmetry of leptons with the narrow
resonance (via $e.g.~K^*$) assumption is
\bea
A^{FB_2}(M_{K^*},z) = \frac{ 8 \pi F_2(M_{K^*},z)}{3 B(M_{K^*},z)}.
\eea
If no new CP
phases are present, $A^{FB_i}(s,z) = FB_i~ \Gamma_i /B(s,z)$.
We can also define several CP asymmetries,
\bea
A_{CP}(s,z) &=& \frac{\bar{\Gamma}_1 - {\Gamma}_1}
                 {\bar{\Gamma}_1 + {\Gamma}_1}, \nn \\
A^{FB_i}_{CP}(s,z) &\equiv& \frac{FB_i[\eta_{CP}\bar{\Gamma}_i - \Gamma_i ]}{\bar{B}(s,z) + B(s,z)},
\label{ACPl}
\eea
where $A_{CP}^{FB_i}$ is the CP asymmetry for each $FB_i$ asymmetry. 
(The CP asymmetry for $FB_2$ was also defined in \cite{CPFB}.) 
Similarly, the time dependent CP asymmetries of $B^0 \rightarrow K^0 \pi^0 l^+
l^- $ are defined after combined with the FB asymmetries,
 \bea
S^{FB_i}_{CP}(s,z) = \frac{2 \eta_{CP} Im \left[  e^{-2i\phi_1}~ FB_{i}~
              \Gamma_{total}\left(C_x^* C_y \to C_x^* \bar{C}_y\right)~
                         \right]}
            { FB_{i} \left[ \eta_{CP}\bar\Gamma_{total} + \Gamma_{total} \right]},
\eea
where $C_x^*C_y \to C_x^* \bar{C}_y$ means
the Wilson coefficients of all $C_x^* C_y$ in $\Gamma_{total}$ is
replaced to the charge conjugated Wilson coefficients like $C_x^* \bar{C}_y$.
If there is no new CP phase in the Wilson coefficients except the CKM phase,
$S_{CP}^{FB_i}$ becomes exactly $\eta_{CP}\sin2\phi_1$ after the cancellation of $FB_i$ in Eq. (39).
However, if there  exists any new CP phase beyond the SM,
the values would change
appropriately. Therefore,
investigating the time-dependent CP asymmetries will
be very important to find hints of new physics.

\section{Forward-Backward asymmetries of leptons}

In Fig. 2, we show those newly defined CP averaged FB asymmetries as  functions of the invariant mass square ($z$)
of dilepton,
where the red (solid) line, the green (dashed) line, the blue (dash-dotted) line
and the purple (dotted) line show the SM case, the SM case with $-C_7$,
the case with $C_7^\prime =|C_7|$, and the case with $-C_7^\prime $, respectively.
\begin{figure}[htbp]
\begin{center}
\begin{minipage}[c]{0.4\textwidth}
{\includegraphics[scale=0.35,angle=-90]{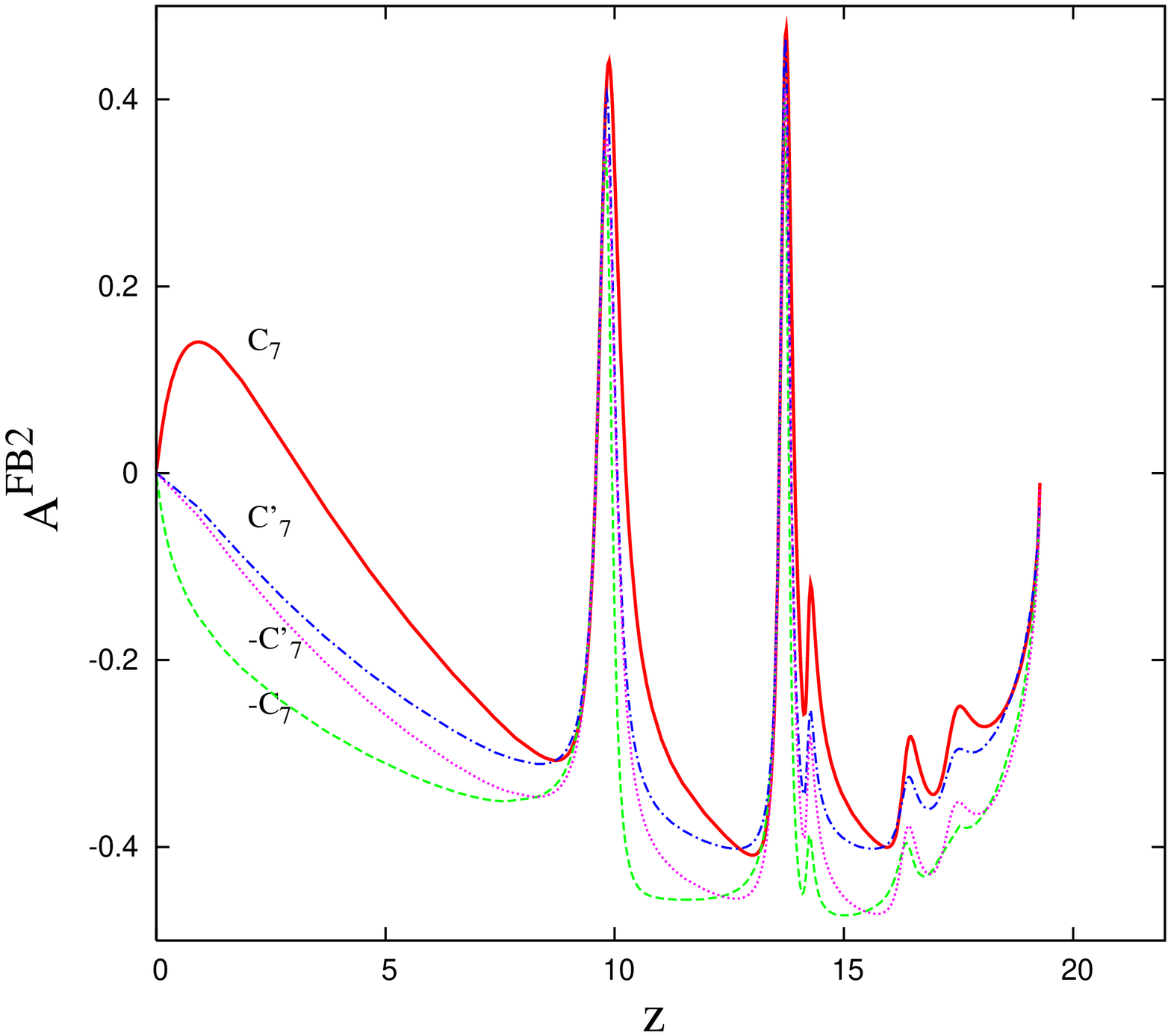}}
\end{minipage}
    \hspace*{10mm}
\begin{minipage}[c]{0.4\textwidth}
{\includegraphics[scale=0.35,angle=-90]{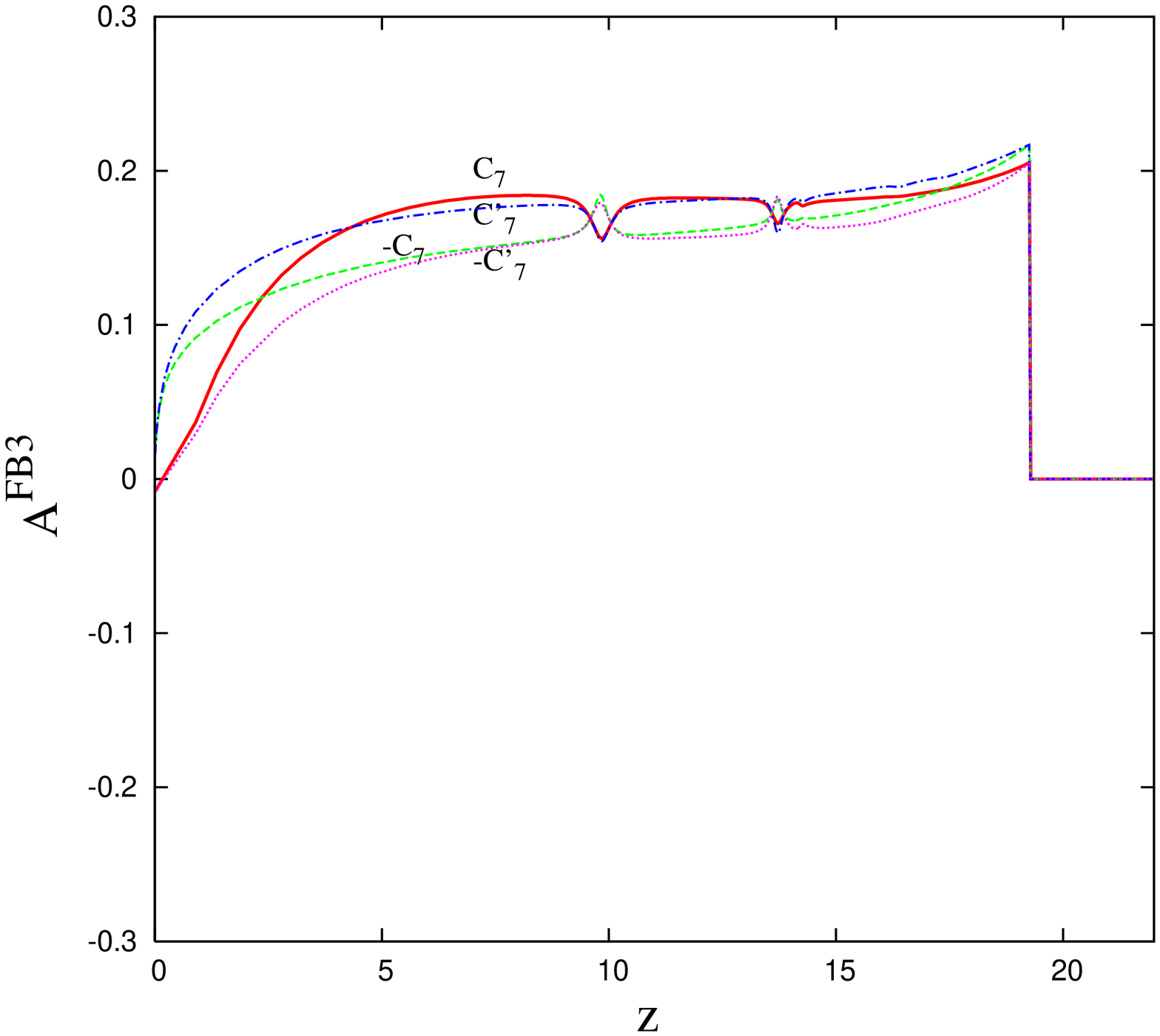}}
\end{minipage} \\[5mm]
\begin{minipage}[c]{0.4\textwidth}
{\includegraphics[scale=0.35,angle=-90]{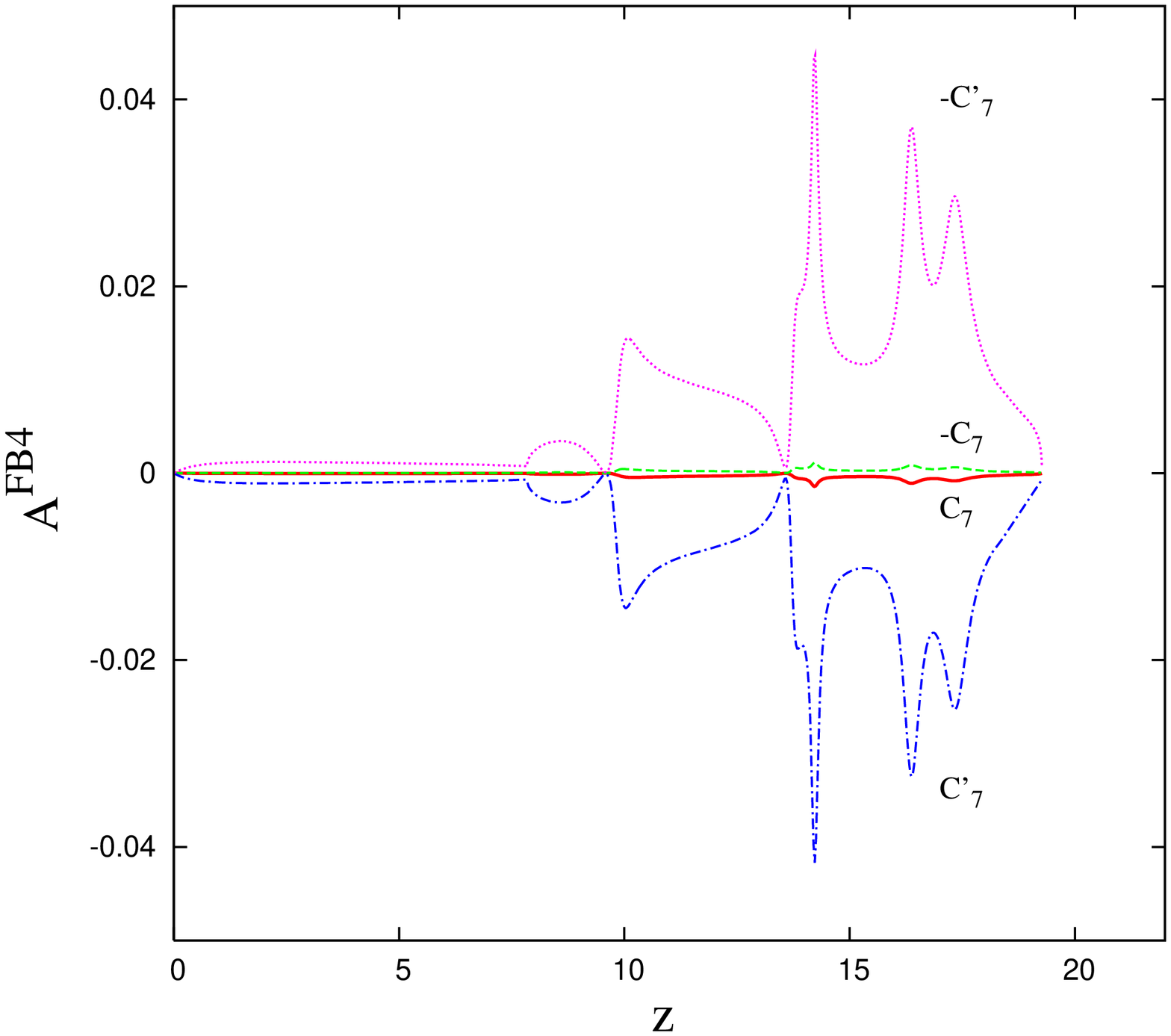}}
\end{minipage}
    \hspace*{10mm}
\begin{minipage}[c]{0.4\textwidth}
{\includegraphics[scale=0.35,angle=-90]{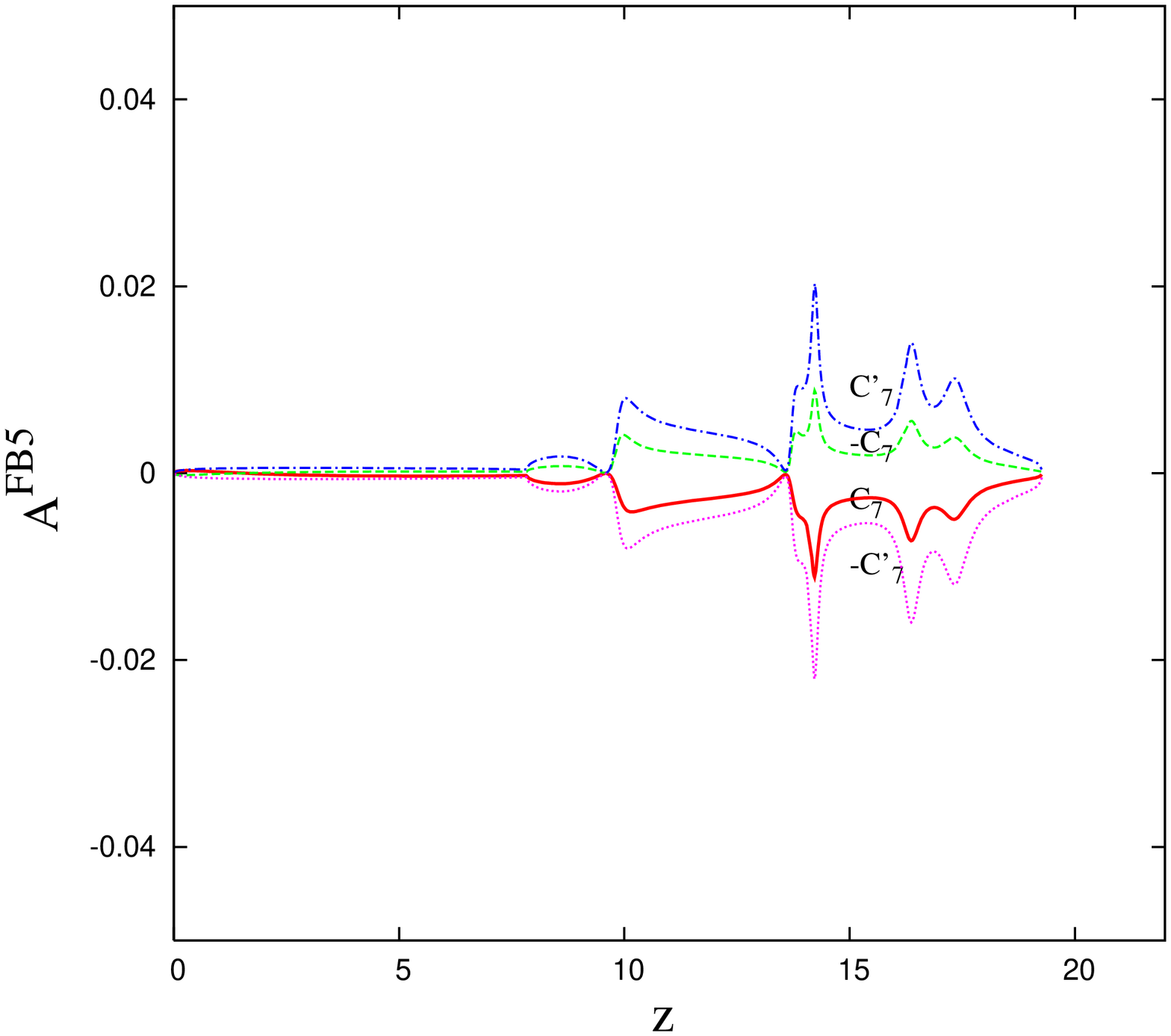}}
\end{minipage} \\[5mm]
\begin{minipage}[c]{0.4\textwidth}
{\includegraphics[scale=0.35,angle=-90]{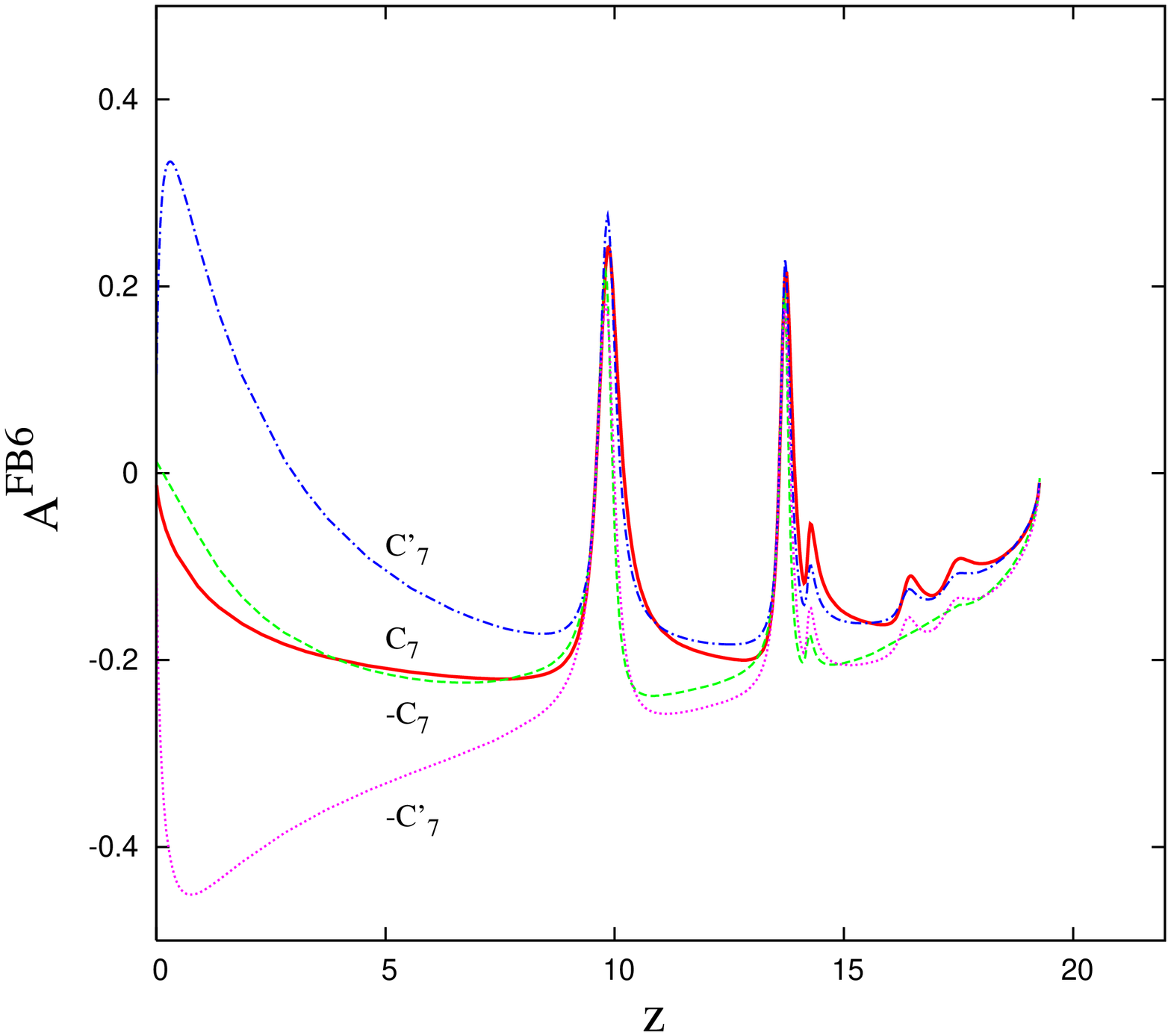}}
\end{minipage}
    \hspace*{10mm}
\begin{minipage}[c]{0.4\textwidth}
{\includegraphics[scale=0.35,angle=-90]{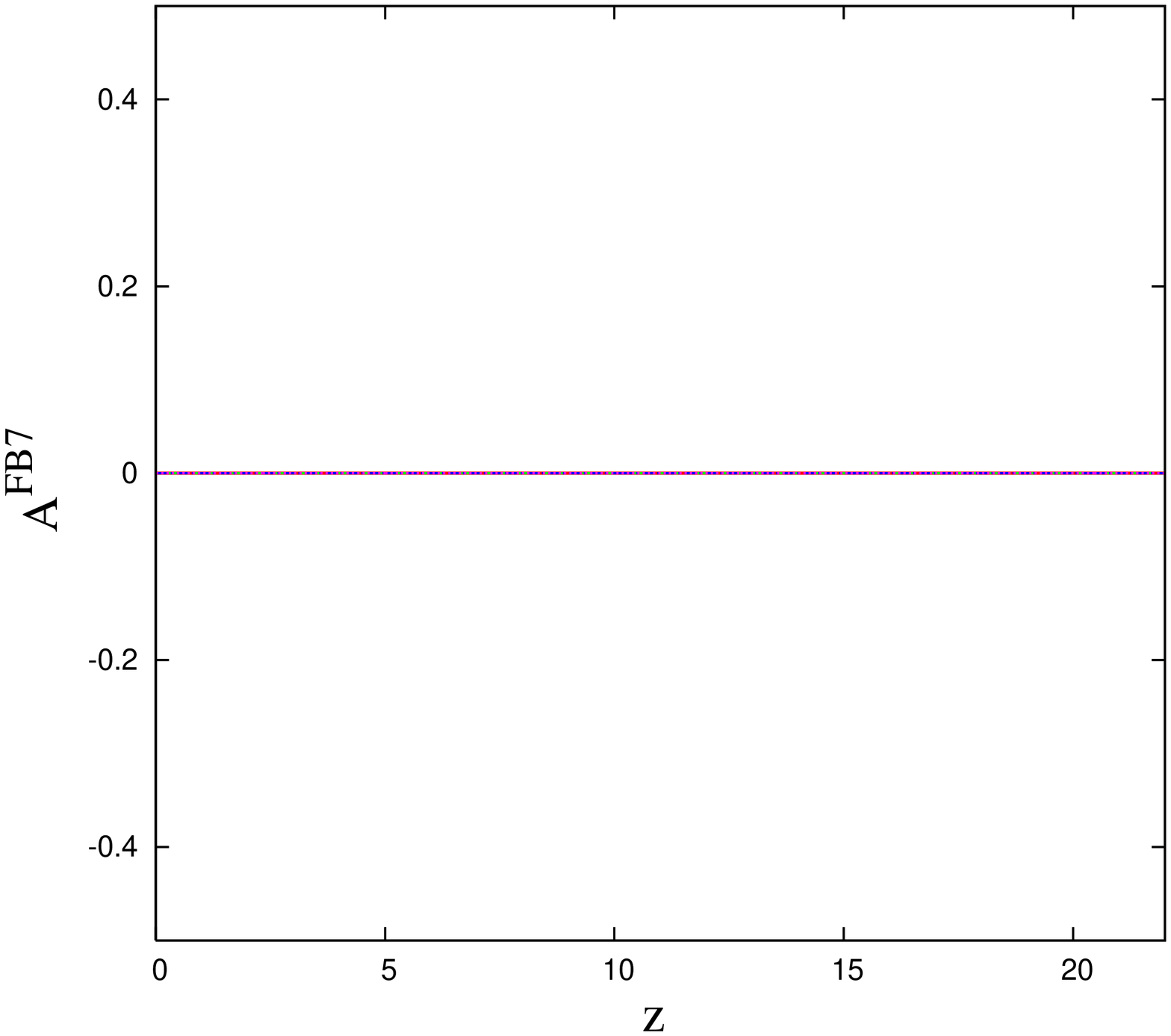}}
\end{minipage}
\end{center}
\caption{FB asymmetries, which defined in Eqs. (\ref{FB2})-(\ref{FB7}), are
plotted, where the solid (red) line shows the SM case, and
the dashed (green) line show
the case with $-C_7$, the dash-dotted (blue) is the pure $C_7^\prime
=|C_7|$
case and the dotted (purple) line is
$-C_7^\prime $ case.
Here, we did not assume any new CP phase and the parameters are
SM predictions except for $C_7$ and $C_7^\prime $.}
\end{figure}
In Fig. 2, we have not assumed any new CP phase and the magnitude of the
parameters are the SM predictions, except for {$C_7^\prime $}.
We note that in the SM $C_7$ and $C_{10}$ are almost real, so that the
origin of CP violation is from the imaginary part of $C_{9}^{eff}$, whose
contributions come from intermediate $\bar{c}c$  bound states.
Due to the absence of strong phase in $C_9^{eff}(z)$ at low dilepton invariant mass region within the SM,
some asymmetries show very strong sensitivity to such extra CP phase, if exists,
which can be an undeniable evidence of new physics with new CP phase.
\begin{figure}[t]
\begin{center}
\begin{minipage}[c]{0.4\textwidth}
{\includegraphics[scale=0.35,angle=-90]{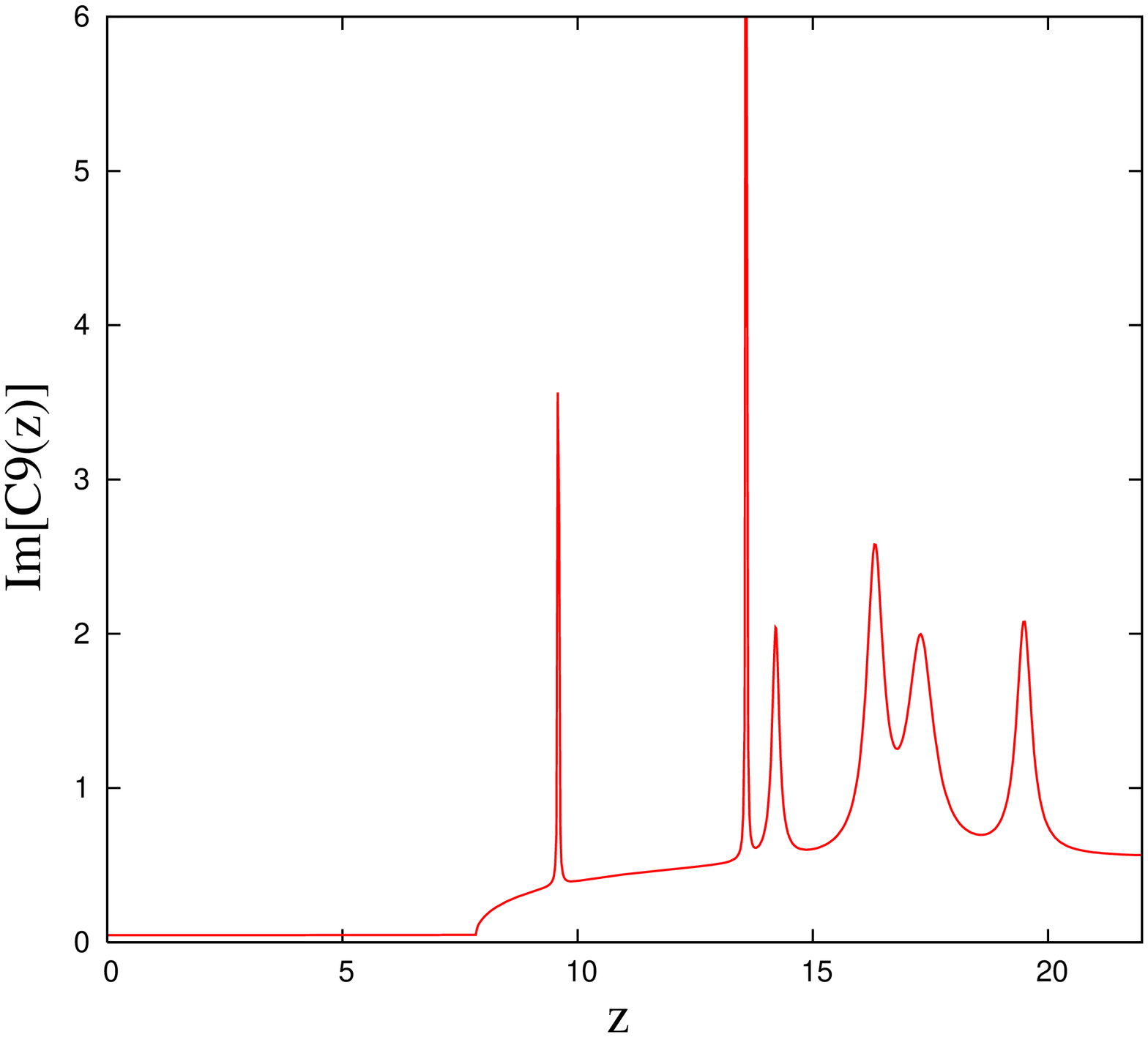}}
\end{minipage}
    \hspace*{10mm}
\begin{minipage}[c]{0.4\textwidth}
{\includegraphics[scale=0.35,angle=-90]{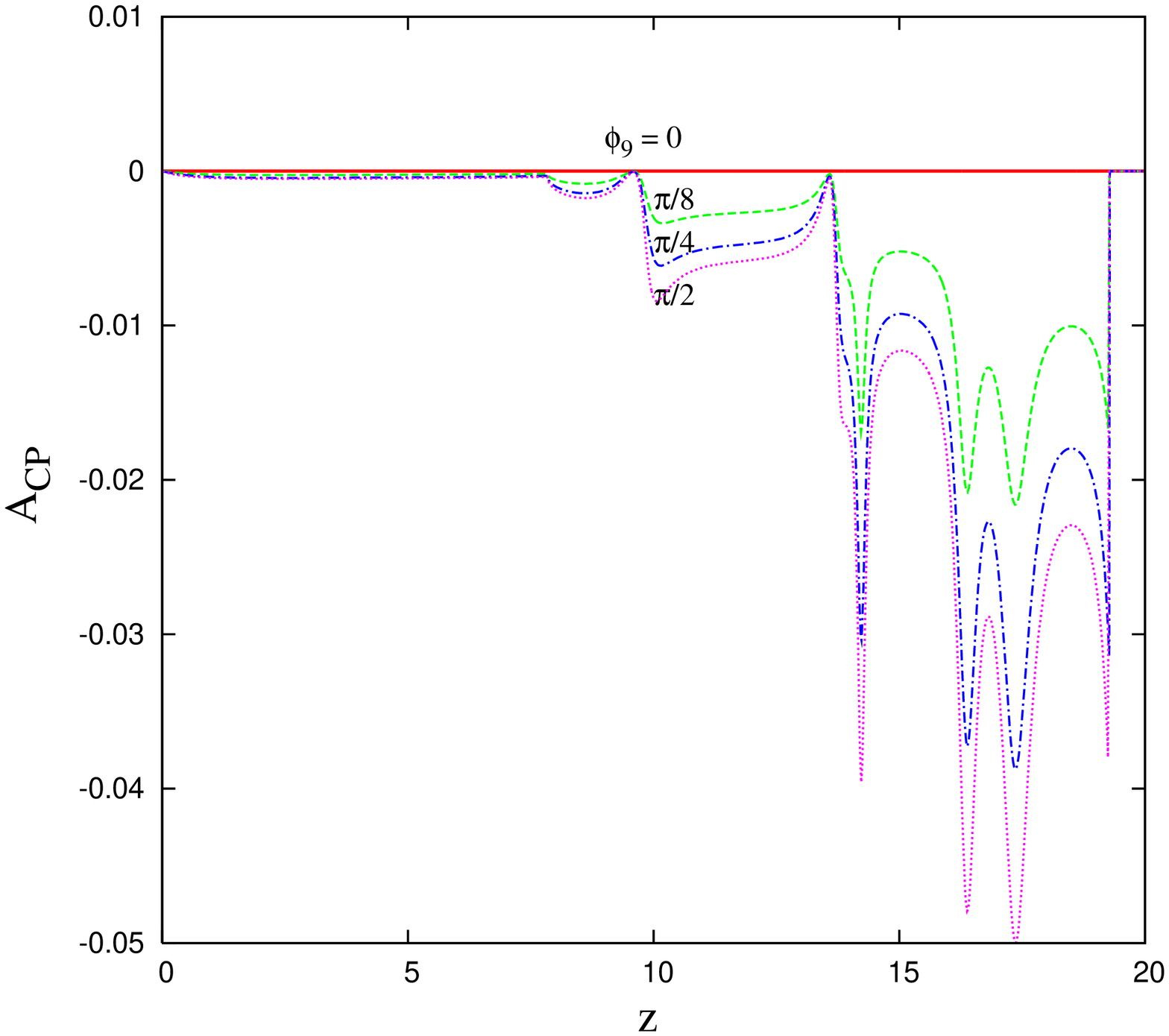}}
\end{minipage}
\end{center}
\label{FIG-IM9}
\caption{ The Imaginary part of $C_9^{eff}$ and the direct CP
asymmetry, $A_{CP}$, for varying the CP phase of $C_9$. }
\end{figure}
Note that $FB_4~ \Gamma_4$ and $FB_5~ \Gamma_5$ can easily extract the imaginary part of
$C_9^{eff}$,   whose contributions within the SM come only from
intermediate $\bar{c}c$
bound stats in high $z$ region.
Hence, those
observables are very sensitive to new phase in low $z$ region.
$FB_7~ \Gamma_7$ does not include $C_9$ and is proportional to $Im[C_7^*C_{10}]$,
so that it can be sensitive to new CP phase in $C_7$ and $C_{10}$.
In the usual case, $C_7$ and $C_{10}$ are almost real (except for
overall factor) so that FB asymmetry for $\Gamma_7$,
$FB_7~{\Gamma_7}$, should be zero.
$A^{FB_2}$ is the usual FB asymmetry for leptons. Therefore,
proving the zero point of the asymmetry,
$A^{FB_2}(z)=0$, in low $z$ region can show the evidence of new physics
contribution.
For $A^{FB_6}$, it is very similar to $A^{FB_2}$ but with the slightly different behavior.

In Fig. 3, we  show the imaginary part of $C_9^{eff}$ \cite{CPFB} and the direct CP asymmetry
$A_{CP}$ as a function of $z$.
One can find that within the SM the direct CP asymmetry $A_{CP}$ in Fig. 3 is quite small
because it is directly proportional to $C_9^{*eff} C_7$ terms, which is
small and also suppressed by $1/z$. In general a CP asymmetry
for modes with both CP odd and CP even is canceling each other, becoming small.
On the other hand,  the CP asymmetry for $FB_i$ may not be so
because they are enhanced by the angular decomposition.

\begin{figure}[htbp]
\begin{center}
\begin{minipage}[c]{0.4\textwidth}
{\includegraphics[scale=0.35,angle=-90]{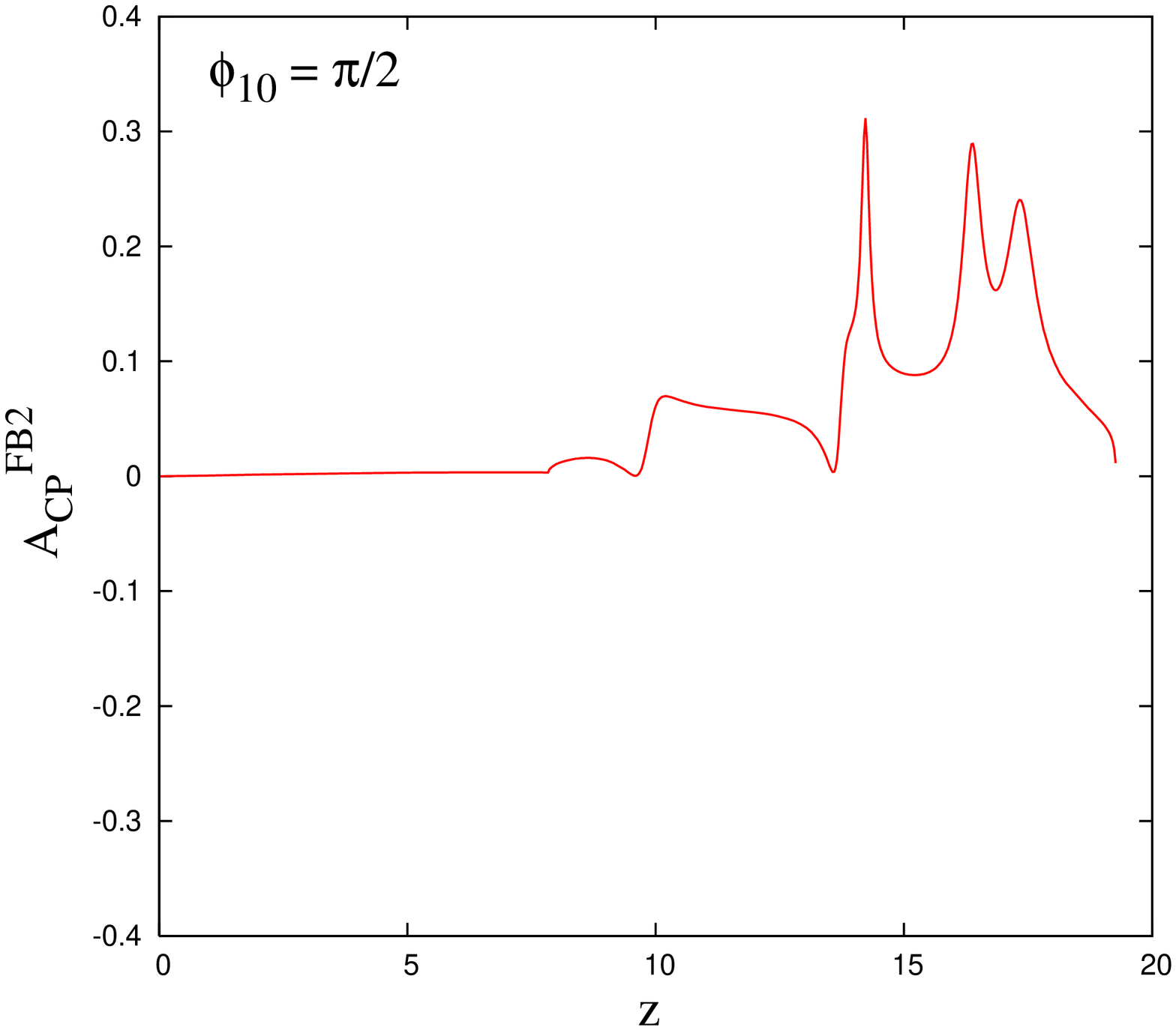}}
\end{minipage}
    \hspace*{10mm}
\begin{minipage}[c]{0.4\textwidth}
{\includegraphics[scale=0.35,angle=-90]{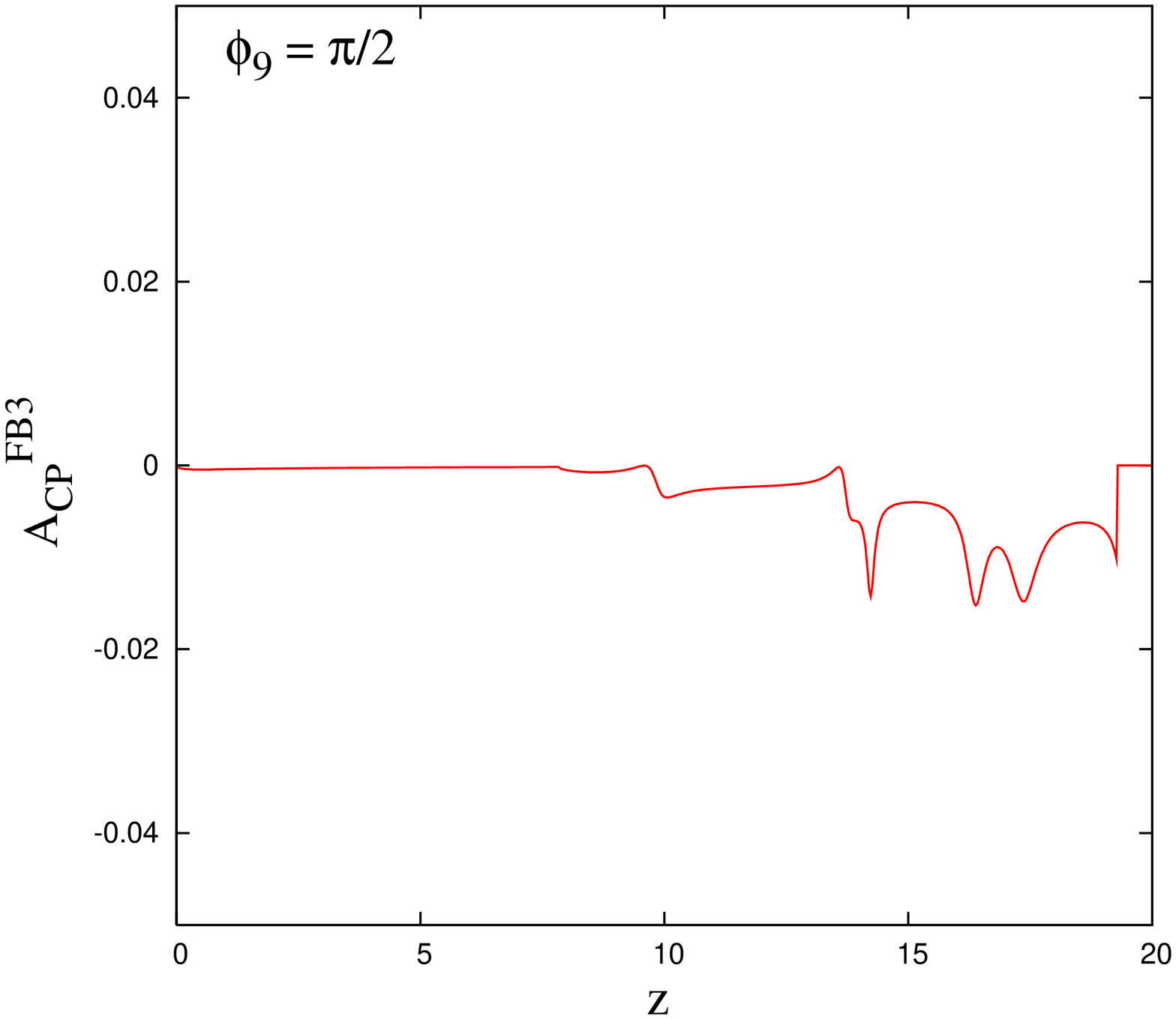}}
\end{minipage} \\[5mm]
\begin{minipage}[c]{0.4\textwidth}
{\includegraphics[scale=0.35,angle=-90]{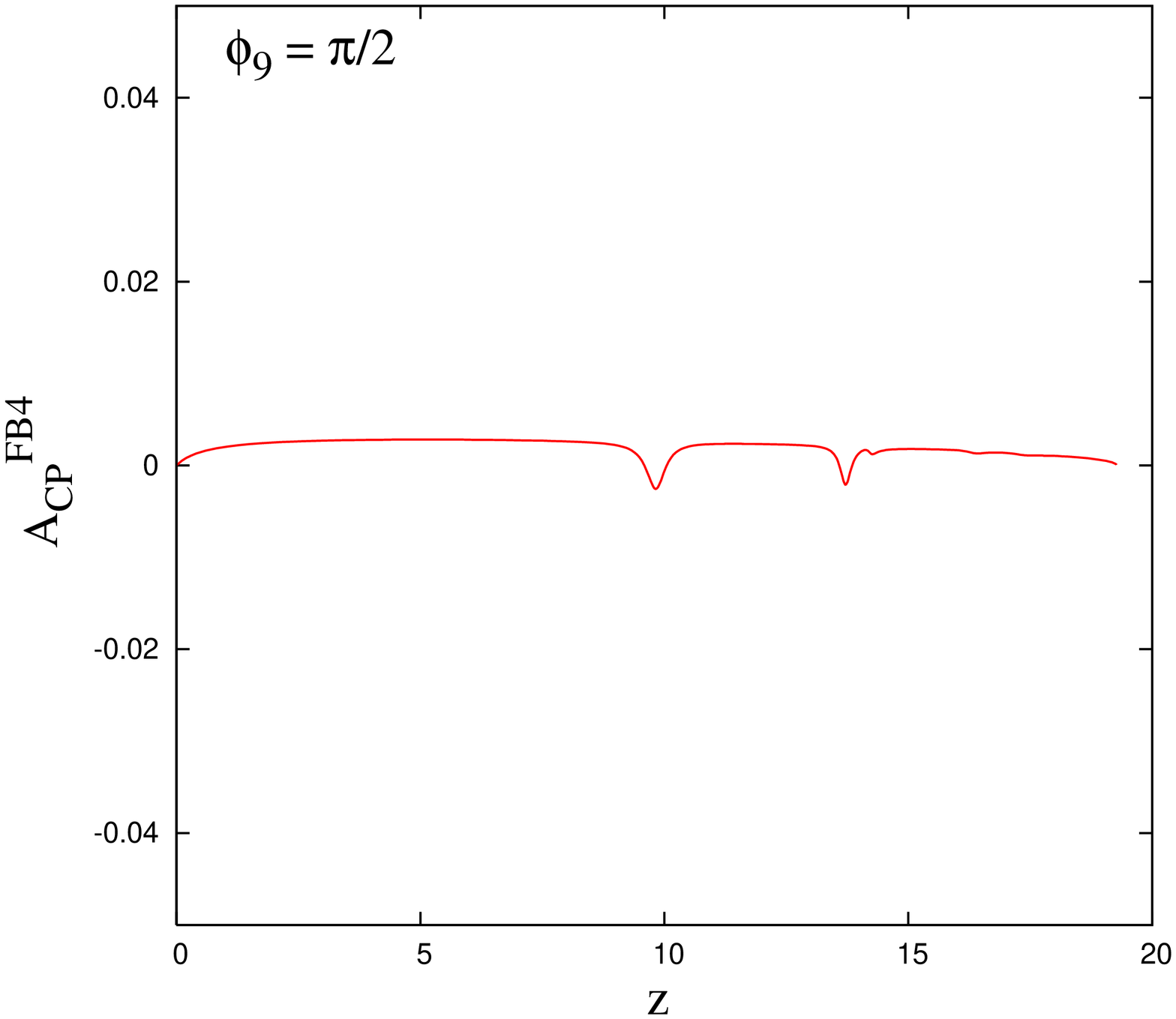}}
\end{minipage}
    \hspace*{10mm}
\begin{minipage}[c]{0.4\textwidth}
{\includegraphics[scale=0.35,angle=-90]{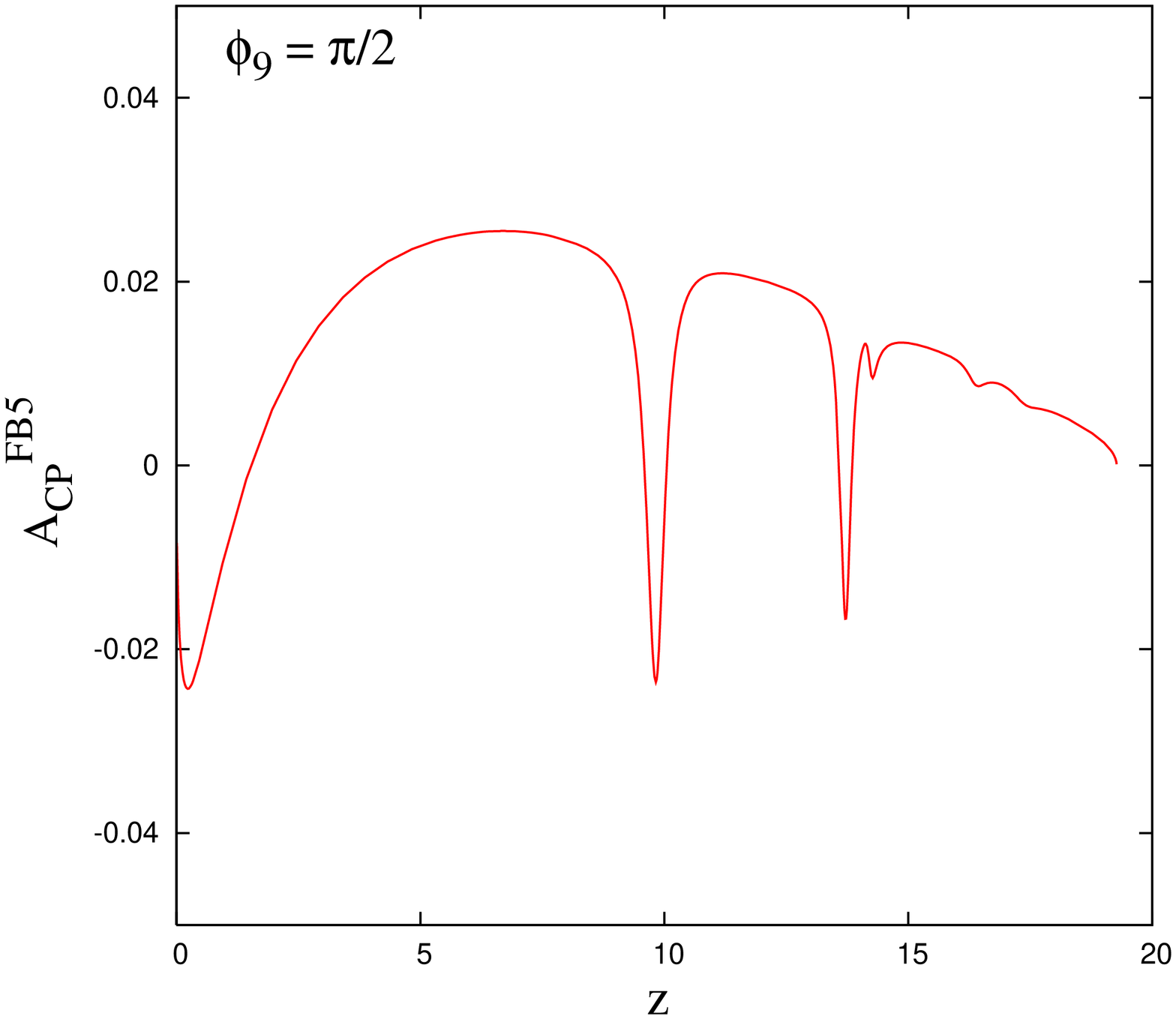}}
\end{minipage} \\[5mm]
\begin{minipage}[c]{0.4\textwidth}
{\includegraphics[scale=0.35,angle=-90]{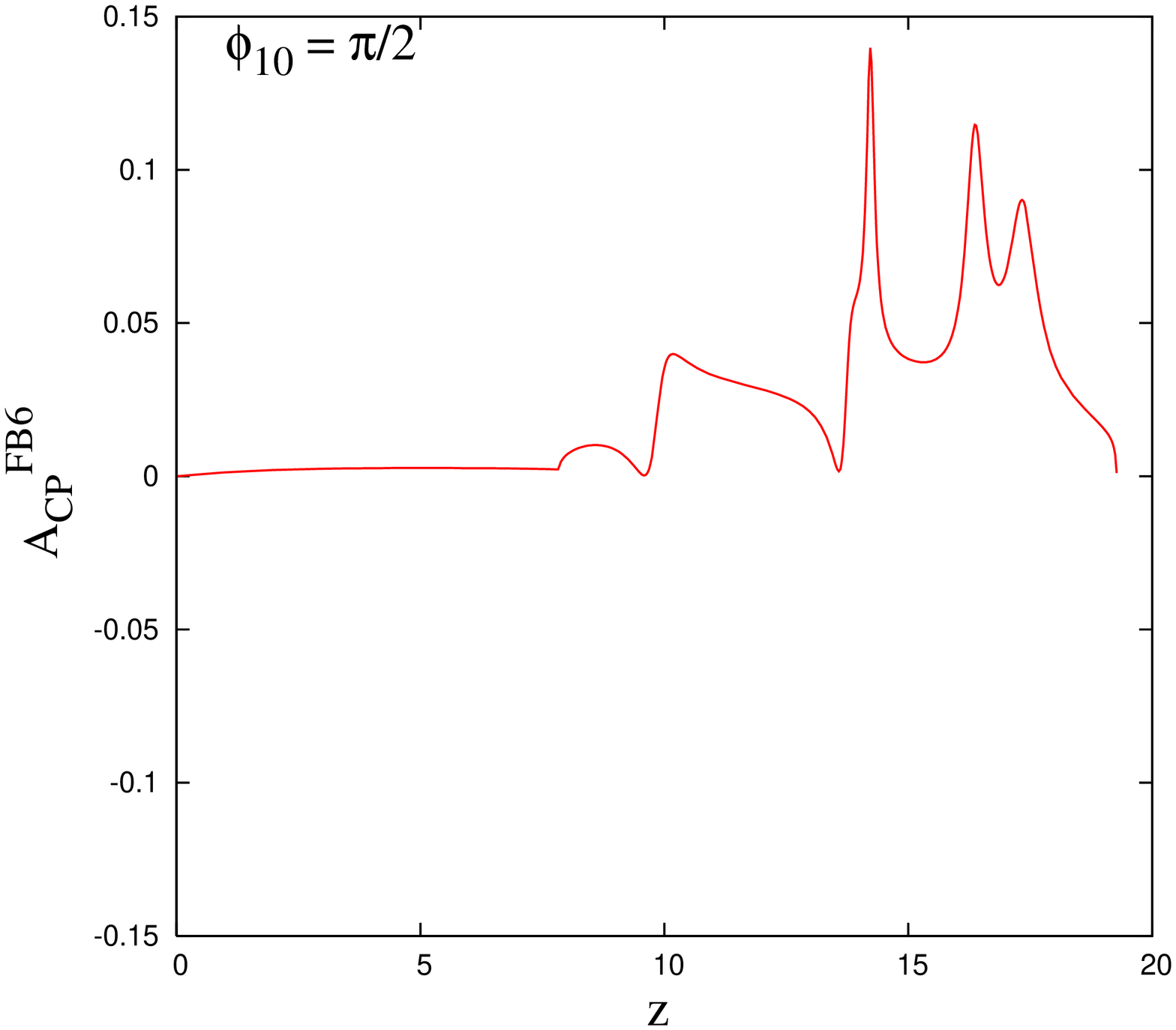}}
\end{minipage}
    \hspace*{10mm}
\begin{minipage}[c]{0.4\textwidth}
{\includegraphics[scale=0.35,angle=-90]{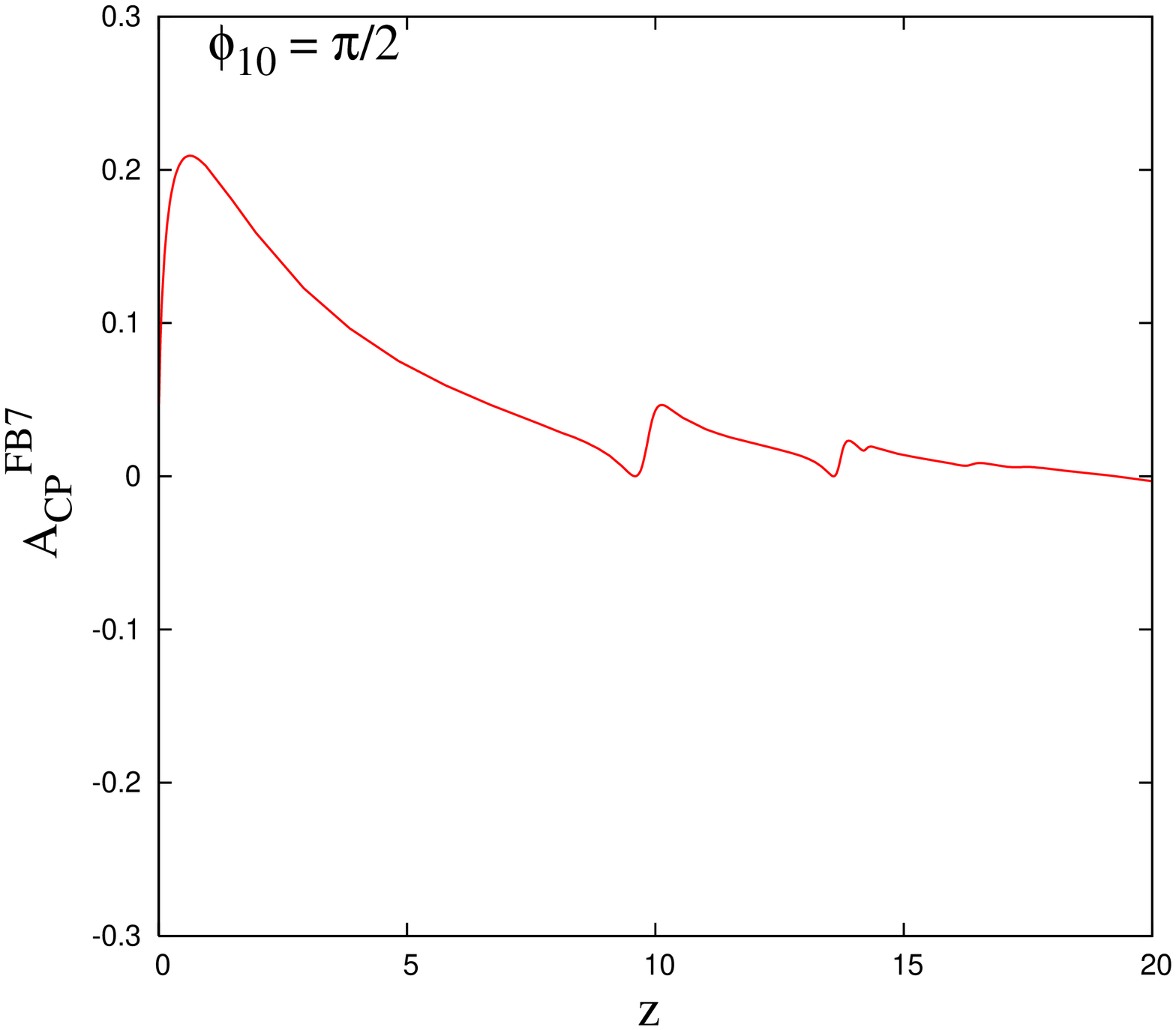}}
\end{minipage} \\[5mm]
\end{center}
\caption{ For $A_{CP}^{FB_2}, A_{CP}^{FB_6}, A_{CP}^{FB_7}$,
we show the case that $C_{10}$ has a pure imaginary CP phase.
For $A_{CP}^{FB_3}, A_{CP}^{FB_4}, A_{CP}^{FB_5}$, we introduced a pure imaginary CP phase in $C_9$. }
\end{figure}

CP asymmetry for each FB asymmetry, $A_{CP}^{FB_i}$, is plotted in Fig 4
as a function of the dilepton's invariant mass.
Here we have introduced new CP phase in $C_9$ and $C_{10}$.
The lines for $A_{CP}^{FB_2}, A_{CP}^{FB_6}, A_{CP}^{FB_7}$ are showing the case that
$C_{10}$ has a pure imaginary CP phase. For $A_{CP}^{FB_3}, A_{CP}^{FB_4}$ and $A_{CP}^{FB_5}$,
we introduced a pure imaginary CP phase in $C_9$.
Please note  a condition to have
CP asymmetry is the existence of  strong phase differences among several contributions.
In the figures of $A_{CP}^{FB_2}$ and $A_{CP}^{FB_6}$,
one can find the dependence of imaginary part of $C_9^{eff}$, where
large CP asymmetries do not appear in low $z$ region. On
the other hand, the figures of $A_{CP}^{FB_5}$ and $A_{CP}^{FB_7}$
show large CP asymmetries in low $z$ region. This is a very interesting
feature. It is because $\Gamma_7$ is proportional to $Im[C_{10}^*C_7]$ so that
the CP asymmetry has to be $Re[C_{10}^*C_7]\sin[\phi_{10}]$, where
$\phi_{10}$ is newly introduced CP phase of $C_{10}$ as $C_{10}e^{i\phi_{10}}$. Hence from these
observables in  low $z$ region, we can extract a few hints about new CP
phase.

\begin{figure}[b]
\begin{center}
\begin{minipage}[c]{0.4\textwidth}
{\includegraphics[scale=0.35,angle=-90]{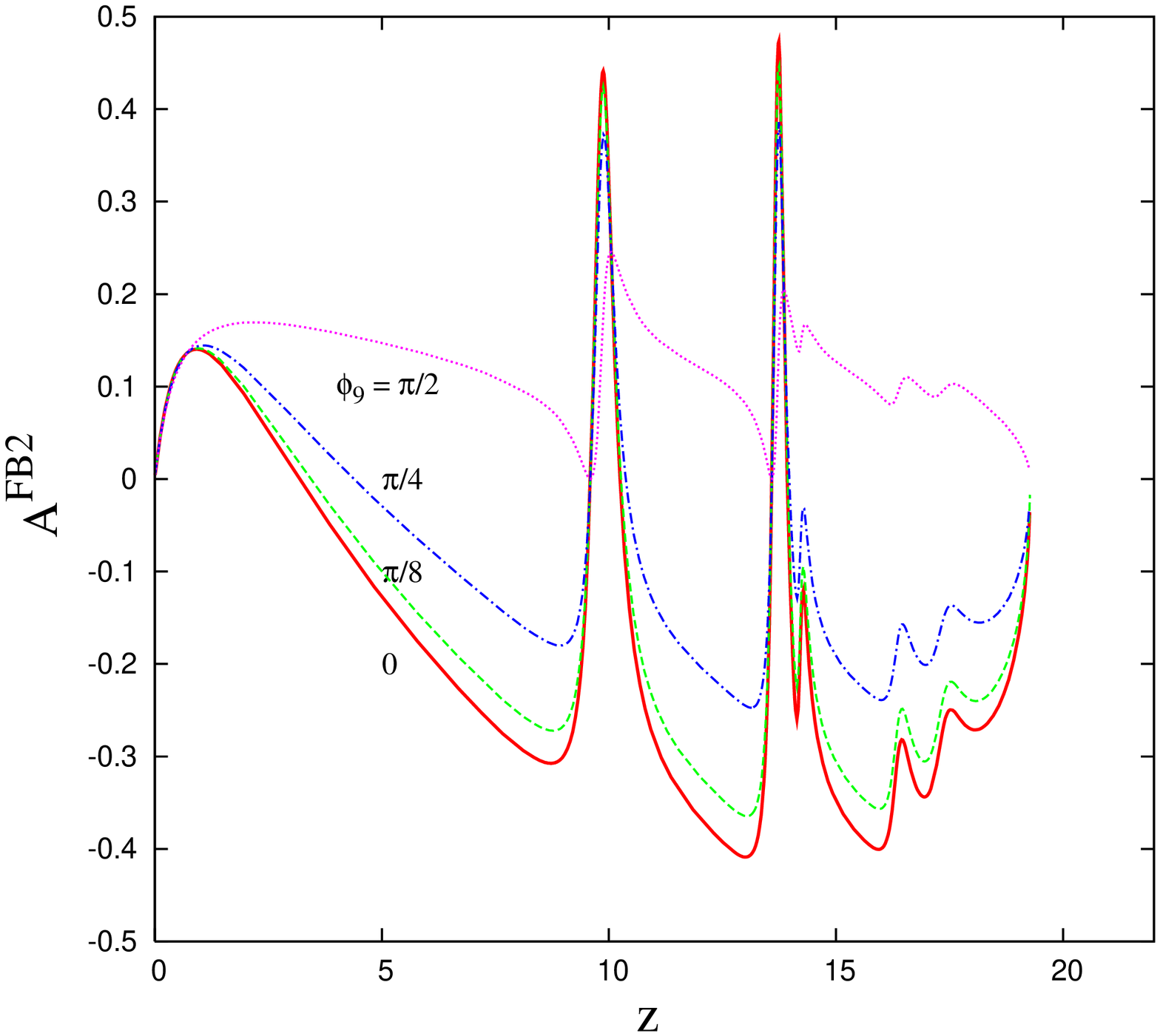}}
\end{minipage}\\[5mm]
\begin{minipage}[c]{0.4\textwidth}
{\includegraphics[scale=0.35,angle=-90]{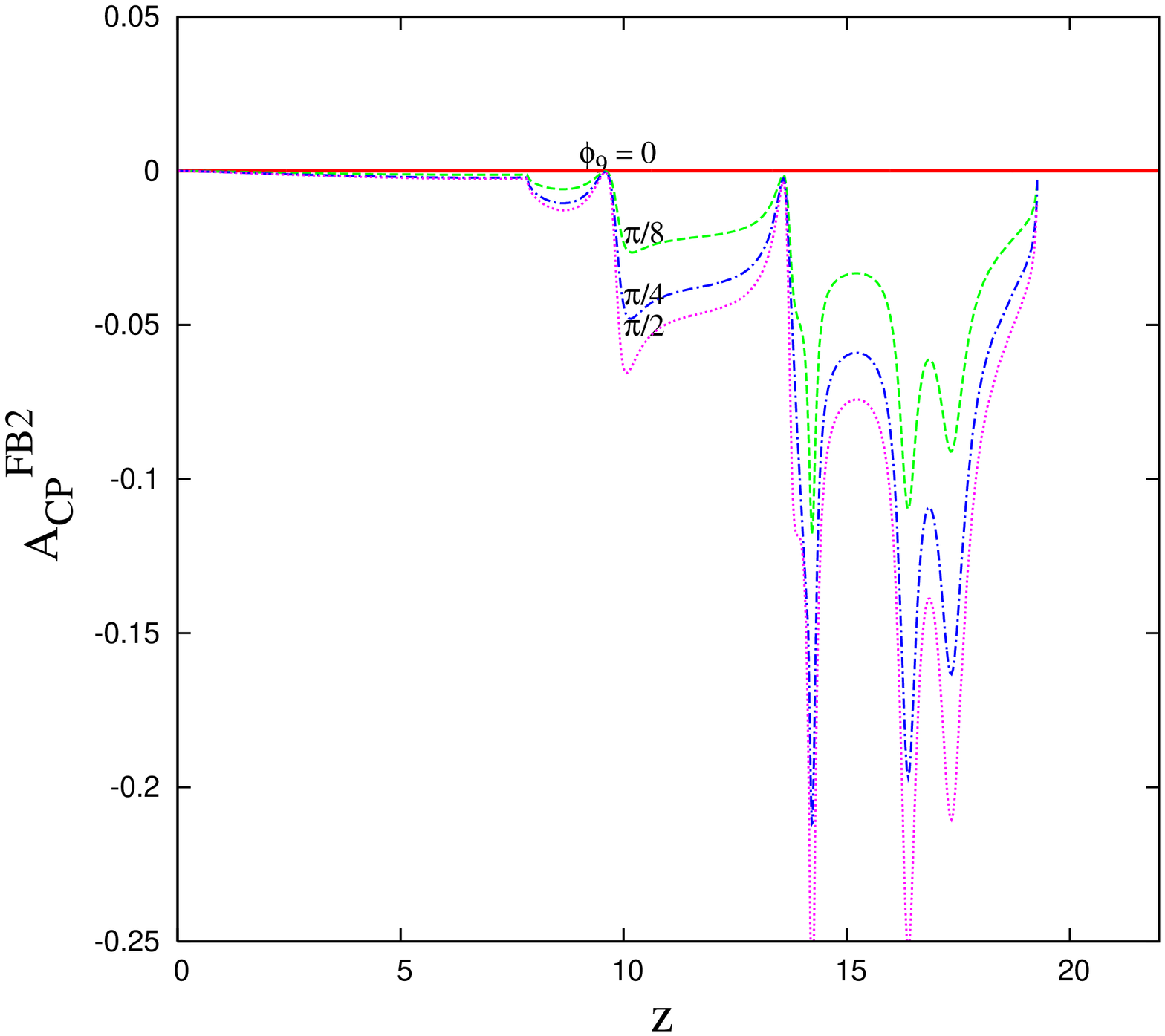}}
\end{minipage}
    \hspace*{10mm}
\begin{minipage}[c]{0.4\textwidth}
{\includegraphics[scale=0.35,angle=-90]{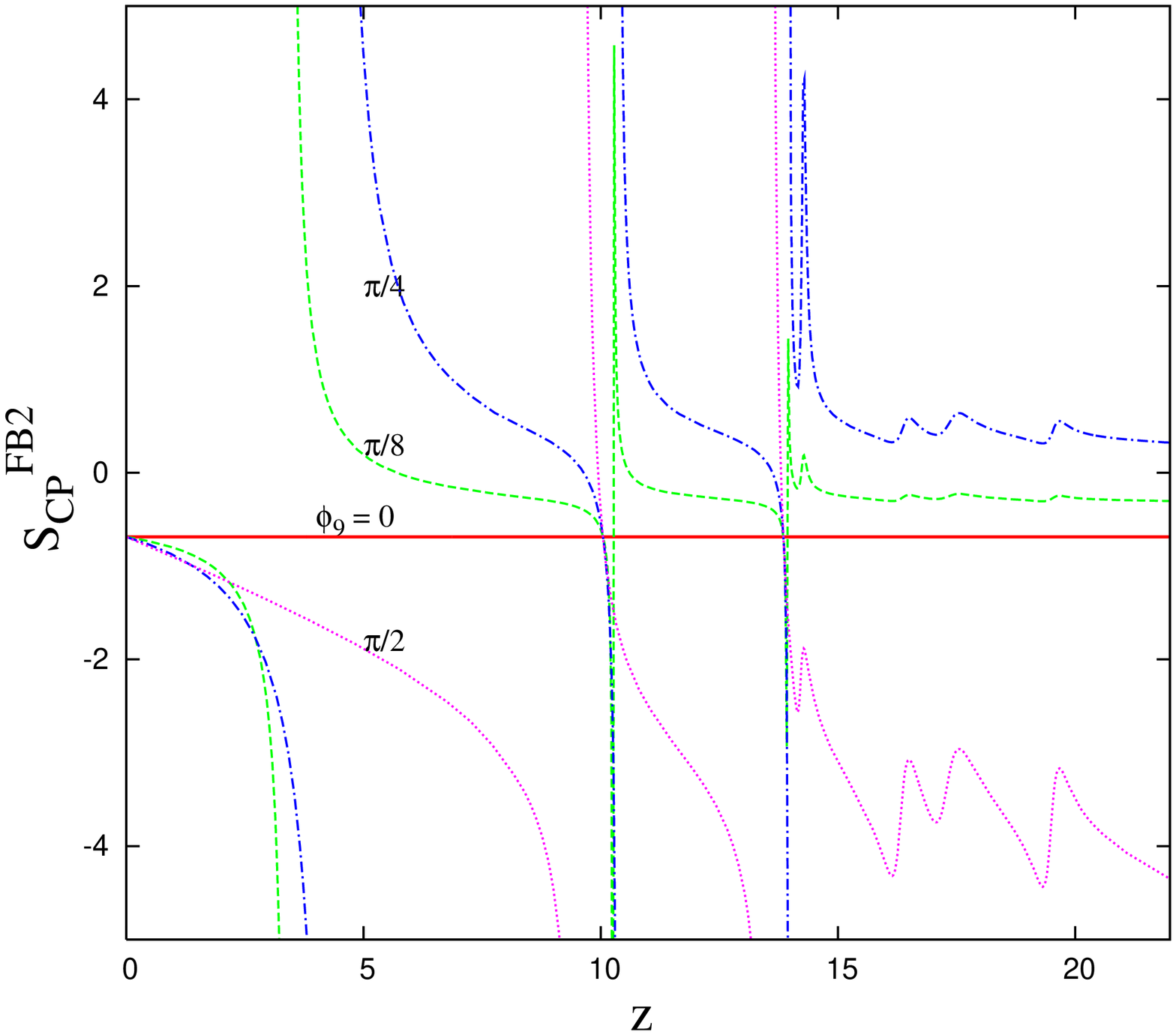}}
\end{minipage}
\label{FIG-EX}
\caption{$A^{FB_2}$, $A_{CP}^{FB_2} $ and $S_{CP}^{FB_2}$
are plotted as functions of $z$, where the new phase of $C_9$ is taken as
$0, \pi/8, \pi/4 $ and $\pi/3 $.}
\end{center}
\end{figure}

In Fig. 5, as an example, $A^{FB_2}$, $A_{CP}^{FB_2} $ and $S_{CP}^{FB_2}$
are plotted as  functions of $z$, where the new CP phase of $C_9$ is taken as
$0, \pi/8, \pi/4 $ and $\pi/3 $.
If there is no new CP phase introduced, $S_{CP}^{FB_i}$ becomes exactly
$\sin2\phi_1$ (red line in the figure). However, if there exists new CP phase, the changes are drastic
with new CP phase.
For the case with scalar and/or tensor type new interactions, the effects will appear
in the branching ratio,  $\Gamma_3$ and $\Gamma_7$.
$E.g.$, if  $C_7$ and $C_{10}$ are real values, $A^{FB_7}$ appears as nonzero value only
with the scalar and/or tensor type new interactions.

\section{Case with Scalar resonance in addition to $K^*$}

In previous section, we have examined the case with a single narrow $K^*$ resonance
limit. Only with a single narrow resonance, we do not have the required
large strong phase difference for direct CP violations.
However, there exist also scalar resonances in the decay mode of $B \to K \pi l^+ l^-$, $e.g.$,
$K^*_0(800)$, $K^*_0(1410)$.
In this section, we consider effects of the interference from
the scalar resonances with the existing vector $K_0^*(892)$ state.
We assume  Wigner type resonance formula for simplicity to
express the effects, even though it is known that this formula cannot
describe the effects precisely.

The matrix element is
\bea
<{\cal M}> = <K,\pi|\{ K^*>< K^*| + S><S| \} {\cal M} |B> ,
\eea
where $S$ expresses a scalar resonance state. If the mass of $S$ is very close
to $K^*(892)$ mass, the cross term between the two resonance states will
make large strong phase difference. To calculate the decay rate, we
use the following parametrization for the hadronic matrix elements,
\bea
<S|\bar{s}\gamma_\mu  b |B> &=& 0 , \\
<S|\bar{s}\gamma_\mu \gamma_5 b |B> &=& \left[ (2q +k)_\mu -
           \frac{m_B^2 - m_0^2}{z} k_\mu \right]F_1(z)
       +\frac{m_B^2 - m_0^2}{z} k_\mu F_0(z), \\
<S|\bar{s}i\sigma_{\mu\nu}k^\nu  b |B> &=& 0, \\
<S|\bar{s}i\sigma_{\mu\nu}k^\nu \gamma_5 b |B> &=& - \frac{1}{m_B +
m_0}
       [ (2q + k)_\mu q^2 - q_\mu (m_B^2 - m_0^2) ] F_T(z), \\
<S|\bar{s}b |B>&=&0, \\[3mm]
<S|\bar{s}\gamma_5 b |B>&=&-\frac{m_B^2-m_{0}^2}{m_B} F_0(z), \\
<S|\bar{s}\sigma_{\mu\nu}k^\nu b |B> &=& 0,
 \eea
where $F_x$ are the form factors. Here we are using the definitions of
Ref. \cite{Ali2}. We also assume that the scalar resonance states as
follow:
\bea
<K\pi|S><S| = m_0 g_{0}\frac{1}{G_0} = - m_0 g_{0}\frac{1}{m_0^2 - s - i m_0 \Gamma_0 },
\eea
where $m_0$ is the mass and $\Gamma_0$ is the decay width of the scalar resonance.
Here we assumed the mass and width of the scalar particle,
$e.g.$, $K_0^*(800)$ \cite{DescotesGenon:2006uk} as,
\bea
 m_0 &=& 0.658~ {\rm GeV},  \nn \\
 \Gamma_0 &=& 0.557~ {\rm GeV}, \nn
\eea
and also assumed that $K^*(800)$ decays  only to $K\pi$.

Using the parameterizations,
the differential decay rate from scalar resonance is
\bea
\Gamma_1^s =
\frac{g_0^2}{|G_0|^2} &\times &[\frac{m_0^2 L^2}{2} \sin^2 \theta_l
   \left\{ \left( |C^{eff}_9(z)- C_9^\prime |^2 + |C_{10} -
           C_{10}^\prime |^2 \right)
           ~(F_1^2) \right.
                   \nn \\
    &+&4 Re\left( (C^{eff}_9(z) - C_9^{\prime *}) (C_7 - C_7^\prime) \right)
           \frac{ m_b }{m_B + m_0} ~(F_1 F_T) \nn \\
    &+&\left.4|C_7 - C_7^\prime|^2 \frac{m_b^2}{(m_B + m_0)^2} ~ |F_T|^2
           \right\} \\
  &+& \left(|C_{AS}|^2 + |C_{AA}|^2\right)
          \frac{2 z m_0^2 (m_B^2 - m_{0}^2)^2}{m_B^2} |F_0|^2 ]. \nn
\eea
And the cross terms with vector $K^*$ resonance contribution are
\bea
\Gamma_2^s &\equiv & F_2^s \cos\theta_K \sin^2\theta_l
                    +F_2^{s\prime }\cos\theta_K\\
&=&  \frac{g_{K\pi} g_0}{|G|^2|G_0|^2} [ \cos\theta_K \sin^2\theta_l \nn \\
  &\times &  \left\{  \left( |C^{eff}_9(z) -C_9^\prime |^2 + |C_{10} -
  C_{10}^\prime |^2 \right)~Re[G G_0^*]
             \left[  \frac{m_0 (m_b+m_{K^*}) L L_0 }{2}
                          ~(A_1 F_1)\right. \right.\nn \\
& & ~~~~~~~~~~~~~~~~~~~~~~~~~~~~~~~~~~~~~~~~~~~~~
   \left.  -\frac{m_0 L^3}{2(m_b + m_{K^*})} ~(A_2 F_1)  \right] \nn \\
& & - 4 |C_7 - C_7^\prime|^2 Re[G G_0^*] \left[ \frac{m_0 m_b^2 L}{2(m_B+m_0) z }
(L^2 - (m_B^2 - m_{K^*}^2 ) L_0 )  ~(T_2 F_T) \right. \nn \\
& &~~~~~~~~~~~~~~~~~~~~~~~~~
    -  \left. \frac{m_0 m_b^2 L}{2 (m_B^2 - m_{K^*}^2 )(m_B+m_0)}
~(T_3 F_T) \right] \nn \\
& & - 4 Re\left( (C^{eff}_9(z) - C_9^\prime )^*(C_7 - C_7^\prime) G  G_0^*
\right)\left[
           \frac{m_0 m_b LL_0 (m_B + m_{K^*})}{4(m_B + m_0)} ~(A_1
           F_T) \right. \nn \\
& & ~~~~~~~~~~~~~~~~~~~~~~~~~~~~~~~~~~~~~~~~~~~~~ -
      \left.     \frac{m_0 m_b L^3 }{4(m_B + m_{K^*})(m_B + m_0)} ~(A_2 F_T) \right]\nn
           \\
& & - 4 Re\left( (C^{eff}_9(z) - C_9^\prime)^*(C_7 - C_7^\prime) G^*  G_0 \right) \left[
            \frac{m_0 m_b L }{4 z}(L^2 -(m_B^2 - m_{K^*}^2) L_0 ) ~(T_2
           F_1) \right. \nn \\
& &  ~~~~~~~~~~~~~~~~~~~~~~~~~~~~~~~~~~~~~~~~~~~~~ +
        \left.    \frac{m_0 m_b L^3 }{4(m_B^2 - m_{K^*}^2)} ~(T_3 F_1) \right]
           \} \\
& & ~~~~~~~ +\cos\theta_K\{ (|C_{AS}|^2 + |C_{AA}|^2 )~Re[G G_0^*]
              \frac{4 z L m_0 m_{K^*} (m_B^2 -m_0^2)}{m_B^2} F_0 A_0 \} ],  \\[5mm]
\Gamma_3^s &\equiv& F_3^s \cos\phi \sin\theta_K \sin\theta_l \nn \\
&=&  \frac{g_{K\pi} g_0}{|G|^2|G_0|^2}  \cos\phi \sin\theta_K
\sin\theta_l \nn \\
&\times& \left\{  -2 \left( Re(C_9^{eff *}(z) C_{10} - C_9^\prime
C_{10}^\prime ) Re(G G_0^*) \right) \frac{m_0  \sqrt{s z}L^2}{(m_B+m_0)}V F_1 \right.
\nn \\
& & \left. - 4Re\left( (C_{10} + C_{10}^\prime )^*(C_7-C_7^\prime) G G_0^*
\right)
   \left[ \frac{m_0 m_b \sqrt{s z } L^2}{2 (m_B + m_{K^*})(m_B+m_0)}~(V
   F_T) \right] \right. \nn \\
& & \left. - 4Re\left( (C_{10} - C_{10}^\prime )^*(C_7+C_7^\prime) G^* G_0
\right)
   \left[ \frac{m_0 m_b \sqrt{s z } L^2}{2 z }~(T_1
   F_1) \right] \right\}, \\[5mm]
\Gamma_4^s &\equiv & F_4^s \sin\phi \sin\theta_K \sin\theta_l \nn \\
  &=&  \frac{g_{K\pi} g_0}{|G|^2|G_0|^2} \sin\phi \sin\theta_K
  \sin\theta_l \nn \\
 &\times &\left\{ - 2 \left(Re(C^{eff}_9(z)^* C_{10} - C_9^{\prime *}
  C_{10}^\prime )Im(G G_0^*)\right) m_0 L
  \sqrt{s z} (m_B + m_{K^*} )~(A_1 F_1) \right.\nn \\
 & & - 4 Im\left( (C_{10} - C_{10}^\prime )^*(C_7-C_7^\prime) [G G_0^*] \right)
      \left[ \frac{m_0 m_b L (m_B + m_{K^*}) \sqrt{s z }}{2 (m_B +
  m_0) } ~(A_1 F_T) \right] \nn \\
 & & + 4 Im\left( (C_{10} - C_{10}^\prime )^*(C_7-C_7^\prime) [G^* G_0] \right)
    \left[\frac{m_0 m_b L (m_B^2 - m_{K^*}^2) \sqrt{s z }}{2 z
  } ~(T_2 F_1) \right]  \\
& &\left. +16 Re((C_{AS}^* C_T + 2 C_{AA}^* C_{TE})G^*G_0)[ \frac{\sqrt{sz}(m_B^2
  -m_0^2) m_0 L}{m_B}T_1 F_0 ] \right\},
\eea
\bea
\Gamma_5^s &\equiv& F_5^s \sin\phi \sin\theta_K \sin2\theta_l \nn \\
&=&  \frac{g_{K\pi} g_0}{|G|^2|G_0|^2}  \sin\phi \sin\theta_K
\sin2\theta_l \nn \\
&\times&\left\{ -\left(|C^{eff}_9(z)|^2 +|C_{10}|^2 -|C_9^\prime |^2 -
     |C_{10}^\prime |^2 \right) Im[G G_0^*]
     \left[ \frac{m_0 L^2\sqrt{s z } }{2 (m_B + m_{K^*})} ~(F_1 V) \right]
        \right. \nn \\
& & - 4 Im\left((C_7+C_7^\prime)^*(C_7-C_7^\prime)G G_0^* \right)
     \frac{m_0 m_b^2 L^2 \sqrt{s z }}{2 z (m_B + m_0)} ~(T_1 F_T) \nn \\
& & - 4 Im( (C_9^{eff}(z)+C_9^\prime )^*(C_7-C_7^\prime) G G_0^* ) \frac{m_0 m_b L^2
     \sqrt{s z } }{4 (m_b + m_{K^*})(m_B + m_0)} ~(V F_T) \nn \\
& & \left.+ 4 Im( (C_9^{eff}(z)-C_9^\prime )^*(C_7+C_7^\prime) G^* G_0 ) \frac{m_0 m_b L^2
     \sqrt{s z } }{4 z } ~(T_1 F_1) \right\}, \\[5mm]
\Gamma_6^s &\equiv& F_6^s \cos\phi \sin\theta_K \sin2\theta_l \nn \\
&=&  \frac{g_{K\pi} g_0}{|G|^2|G_0|^2}  \cos\phi \sin\theta_K
\sin2\theta_l \nn \\
&\times&\left\{ \left(|C^{eff}_9(z) - C_9^\prime |^2 +|C_{10} -
     C_{10}^\prime |^2\right) Re[G G_0^*]
     \left[\frac{m_0 L\sqrt{s z } (m_B+m_{K^*})}{2} ~(A_1 F_1) \right]
        \right. \nn \\
& & +4 \left|(C_7-C_7^\prime)\right|^2 Re[G G_0^*]
     \frac{m_0 m_b^2 L \sqrt{s z }(m_B^2 -m_{K^*}^2)}{2 z (m_B+m_0)}
                           ~(T_2 F_T) \nn \\
& & - 4 Re( (C_9^{eff}(z)- C_9^\prime )^*(C_7-C_7^\prime) G G_0^* ) \frac{m_0 m_b L
     \sqrt{s z } (m_B+m_{K^*})}{4 (m_B + m_0)} ~(A_1 F_T) \nn \\
& & \left.+4 Re( (C_9^{eff}(z)-C_9^\prime )^*(C_7-C_7^\prime) G^* G_0 ) \frac{m_0 m_b L
     \sqrt{s z } (m_B^2 -m_{K^*}^2)}{4 z } ~(T_2 F_1) \right\}, \\[5mm]
\Gamma_7^s &\equiv& F_7^s \cos^2\theta_K \cos\theta_l \nn \\
&=&  \frac{g_{K\pi} g_0 L_0}{|G|^2|G_0|^2} \cos^2\theta_K
\cos\theta_l \nn \\
&\times&\left\{  8 Im((C_{AA}^* C_T - 2 C_{AS}^* C_{TE})GG_0^*)
          \frac{ m_0  (m_B^2 - m_0^2)}{m_B} (L_0 - m_B^2 + m_{K^*}^2 + z )
          T_1 F_0 \right\}, \\[5mm]
\Gamma_8^s &\equiv& F_8^s \cos\phi \sin2\theta_K \sin\theta_l \nn \\
&=&  \frac{g_{K\pi} g_0 \sqrt{s z}}{|G|^2|G_0|^2} \cos\phi \sin2\theta_K \sin\theta_l
              \nn \\
&\times&\left\{  8 Im((C_{AA}^* C_T - 2 C_{AS}^* C_{TE})GG_0^*)
          \frac{ m_0 (m_B^2 - m_0^2)}{m_B} (L_0 - m_B^2 + m_{K^*}^2 + z )
          T_1 F_0 \right\}.
\eea

Note that $\Gamma_7^s $ and $\Gamma_8^s $ have the same angular
distributions as $\Gamma_2 $ and $\Gamma_6 $, respectively, and therefore,
their contributions can be extracted by $FB_2$ and $FB_6$
integration operators. All the other $\Gamma_i^s $'s are independent
of previously defined $FB_i$ of Eqs. (\ref{FB2})-(\ref{FB7}). To
extract these contributions, we need new definitions of FB asymmetries
from new integration operators, $FB_i^s$. Namely, these
contributions  appear only in the case with the scalar resonance
effects. If any one of the following type $FB_i^s$ asymmetries
appears, it can be a strong evidence for the scalar resonance contributions. The
new operators $FB_i^s$  are defined as follow:
\bea
FB_{2}^s~ \Gamma_{total}&=&\int_0^{2\pi }d\phi \int_0^{\pi}\sin\theta_l d\theta_l
           \left(\int^{\frac{\pi}{2}}_{0}
               - \int^{\pi}_{\frac{\pi}{2}}\right) \sin\theta_K d\theta_K
           \Gamma_{2}^s = \frac{8 \pi
             F_2^s}{3} + 4\pi F_2^\prime ,
\label{FBs2}
\\
FB_{3}^s~ \Gamma_{total}^s &=&
             \left(\int_{-\frac{\pi}{2}}^{\frac{\pi}{2}}
                    - \int^{\frac{3\pi}{2}}_{\frac{\pi}{2}}\right)d\phi
             \int_{0}^{{\pi}}
                            \sin\theta_K d\theta_K
                \int^{\pi}_{0}
                            \sin\theta_l d\theta_l
              \Gamma_{3}^s = \pi^2
             F_3^s,\\
FB_{4}^s~ \Gamma_{total} &=& \left(\int_{0}^{{\pi}}
                    - \int^{{2\pi}}_{{\pi}}\right)d\phi
               \int_0^\pi
                 \sin\theta_K d\theta_K \sin\theta_l d\theta_l
              \Gamma_{4}^s  = \pi^2 F_4^s, \\
FB_{5}^s~ \Gamma_{total} &=& \left(\int_{0}^{{\pi}}
                    - \int^{2\pi}_{\pi}\right)d\phi
             \left(\int_{0}^{\frac{\pi}{2}}
                    - \int_{\frac{\pi}{2}}^\pi \right)
                            \sin\theta_l d\theta_l
               \int^{\pi}
                            \sin\theta_K d\theta_K
              \Gamma_{5}^s = \frac{8 \pi F_5^s}{3}, \\
FB_{6}^s~ \Gamma_{total}&=&
             \left(\int_{-\frac{\pi}{2}}^{\frac{\pi}{2}}
                    - \int^{\frac{3\pi}{2}}_{\frac{\pi}{2}}\right)d\phi
             \left(\int_{0}^{\frac{\pi}{2}}
                    - \int_{\frac{\pi}{2}}^\pi \right)
                            \sin\theta_l d\theta_l
              \int^{\pi}_{0}
                            \sin\theta_K d\theta_K
              \Gamma_{6}^s = \frac{8 \pi F_6^s}{3}.
\label{FBs6}
\eea
Also note that $FB_2^s~ \Gamma_2^s$ is the FB asymmetry for $K$ meson (or $\pi$). Similarly $FB^s_3~ \Gamma_3^s$ and
$FB^s_4~ \Gamma_4^s$ are the asymmetries for the angle $\phi$ between two decay
planes. $FB^s_5~ \Gamma_5^s$ and $FB^s_6~ \Gamma_6^s$ are defined as the double FB
asymmetries.
The CP averaged FB asymmetries are defined as
\bea
A^{FB_i^s}(s,z) = \frac{FB_i^s[ \eta_{cp} \bar{\Gamma}_i^s + \Gamma_i^s
]}{\bar{B}(s,z) + B(s,z)},
\eea
where $\eta_{CP} = +1$ for CP even case and $-1$ for CP odd.
We can also define several CP asymmetries,
\bea
A^{FB_i^s}_{CP}(s,z) &\equiv& \frac{FB_i^s[\eta_{CP}\bar{\Gamma}_i^s - \Gamma_i^s ]}{\bar{B}(s,z) + B(s,z)}.
\label{ACPls}
\eea

In Fig. 6, we plot the asymmetries defined in
Eqs. (\ref{FBs2})-(\ref{FBs6}). Here we have not assumed any new CP phases.
The red (solid) curve is the case of the SM with the scalar resonance.
The FB asymmetry of $K(\pi )$ meson, $A^{FB^s_2}$, can be relatively
large but the other asymmetries cannot be so large without any new physics CP phases. $A^{FB_5^s}$ and
$A^{FB_6^s}$ are actually tiny because they are extracted only by the
double asymmetries.
\begin{figure}[htbp]
\begin{center}
\begin{minipage}[c]{0.4\textwidth}
{\includegraphics[scale=0.35,angle=-90]{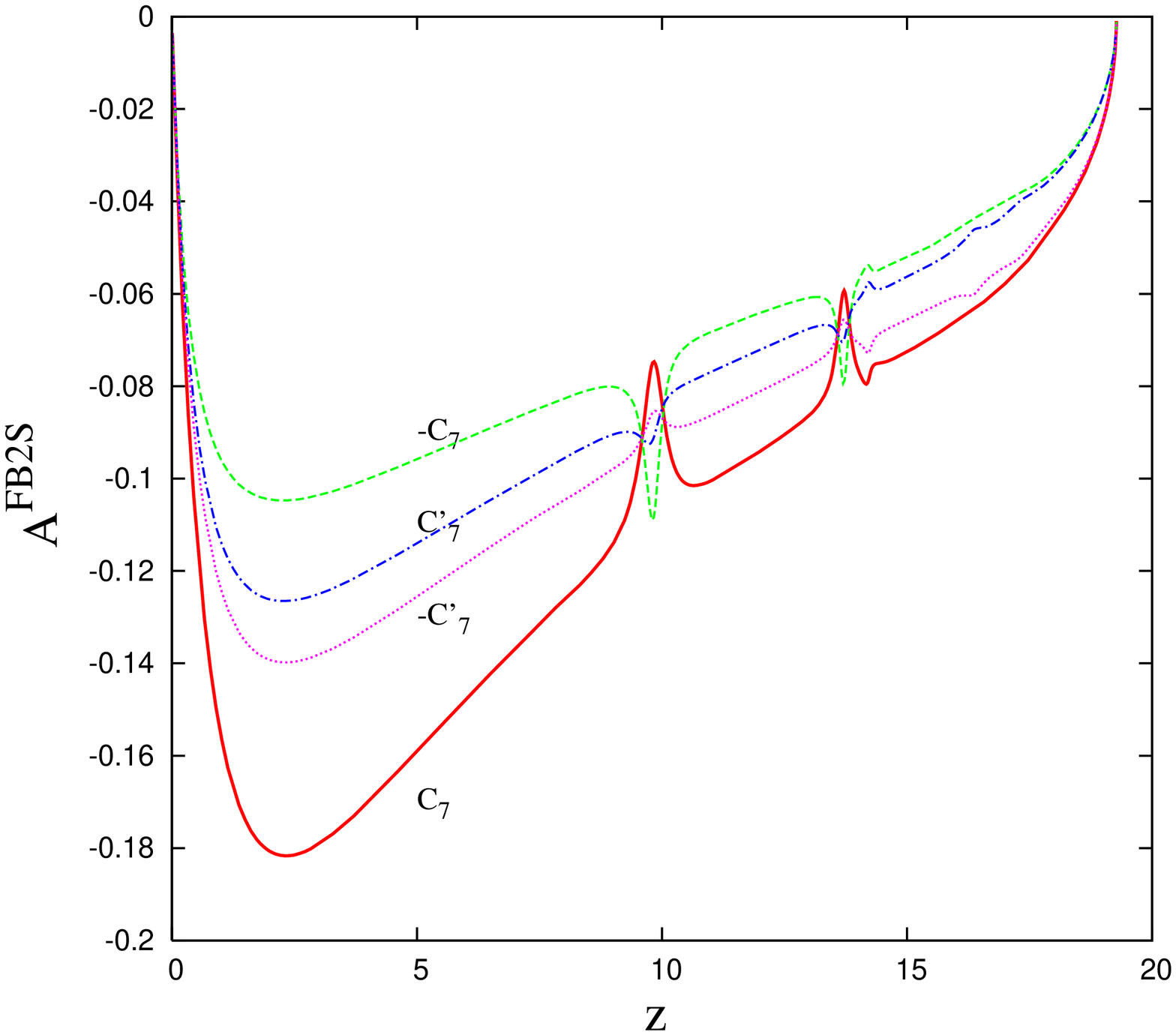}}
\end{minipage}
    \hspace*{10mm}
\begin{minipage}[c]{0.4\textwidth}
{\includegraphics[scale=0.35,angle=-90]{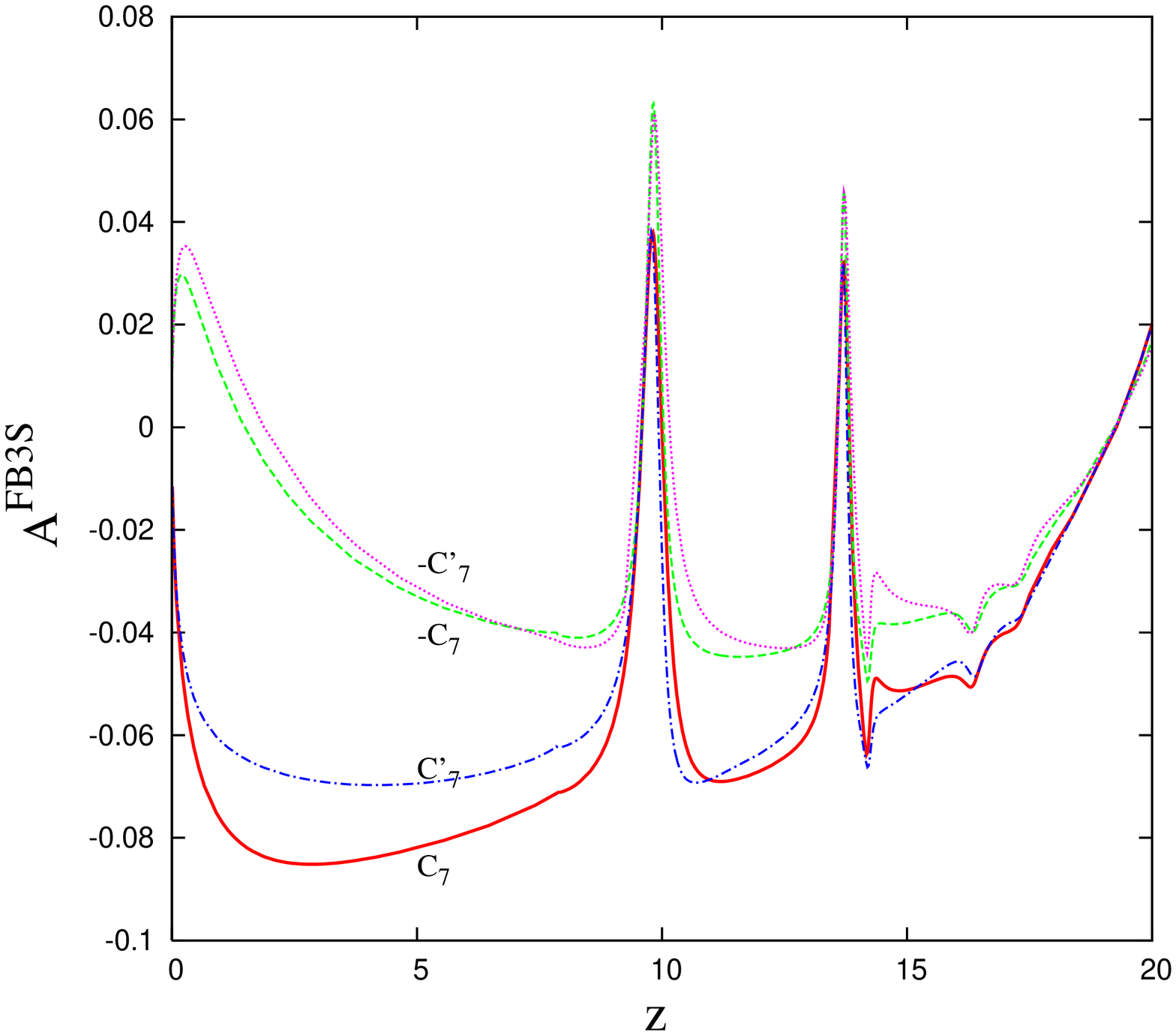}}
\end{minipage} \\[5mm]
\begin{minipage}[c]{0.4\textwidth}
{\includegraphics[scale=0.35,angle=-90]{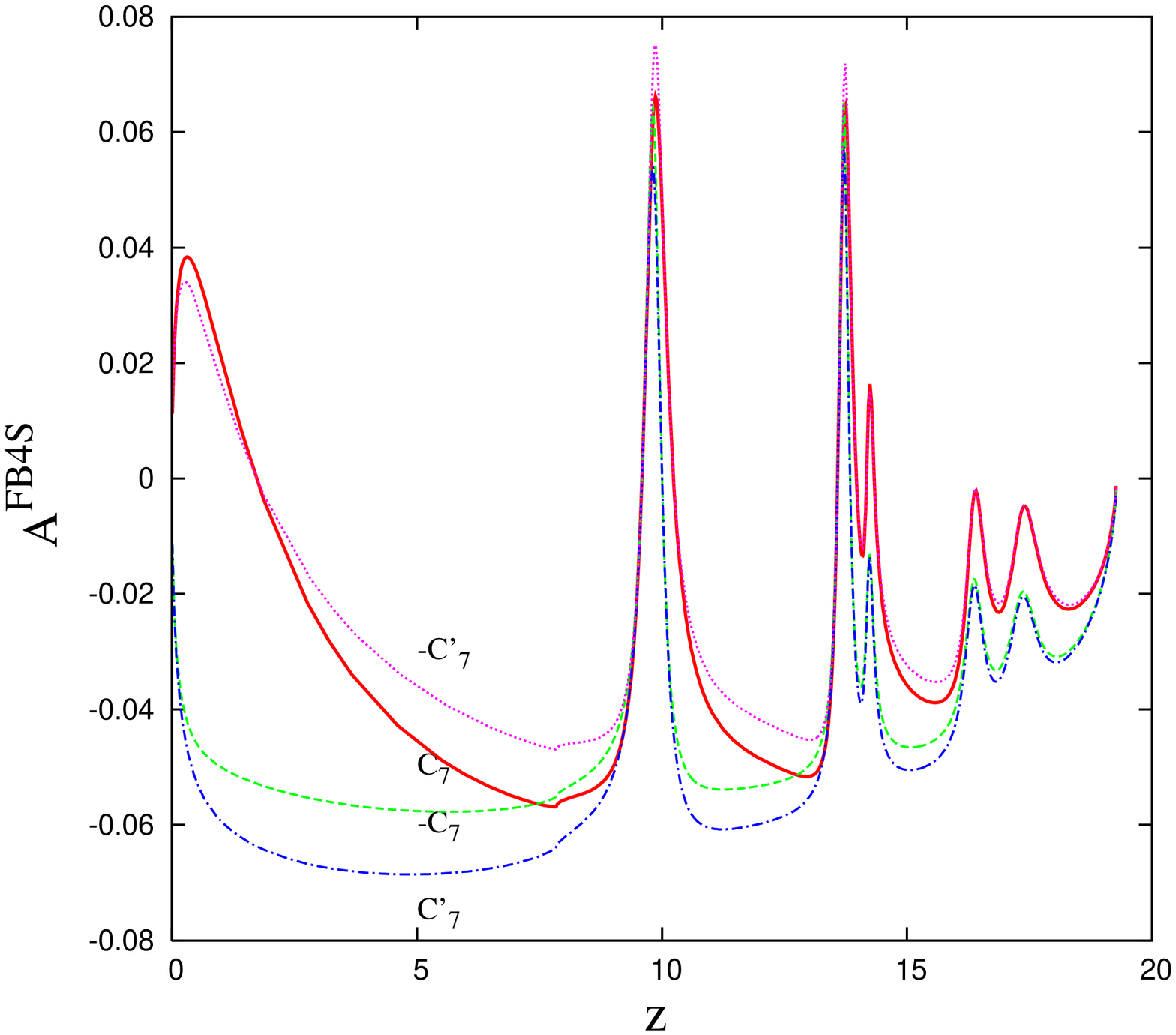}}
\end{minipage}
    \hspace*{10mm}
\begin{minipage}[c]{0.4\textwidth}
{\includegraphics[scale=0.35,angle=-90]{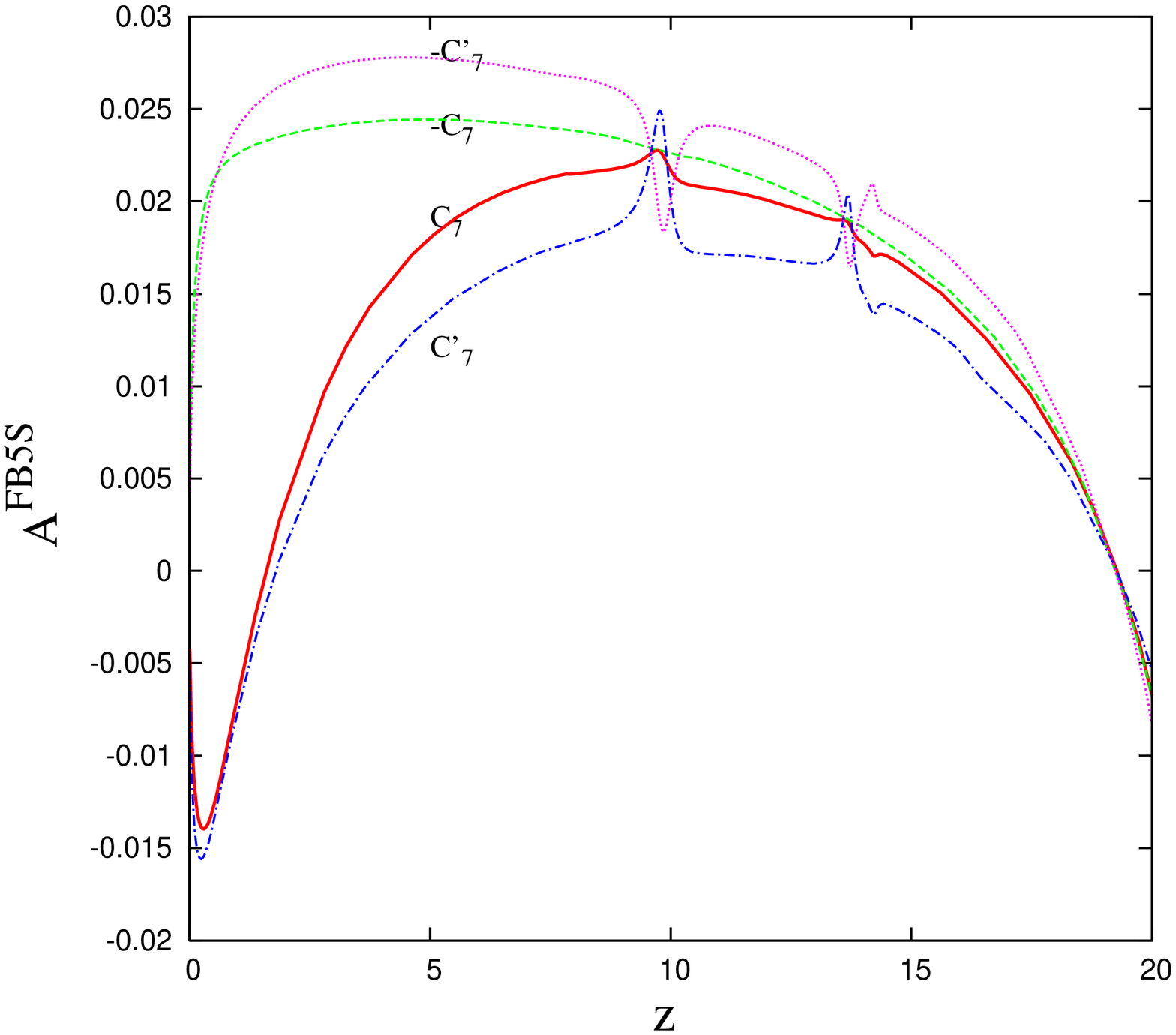}}
\end{minipage} \\[5mm]
\begin{minipage}[c]{0.4\textwidth}
{\includegraphics[scale=0.35,angle=-90]{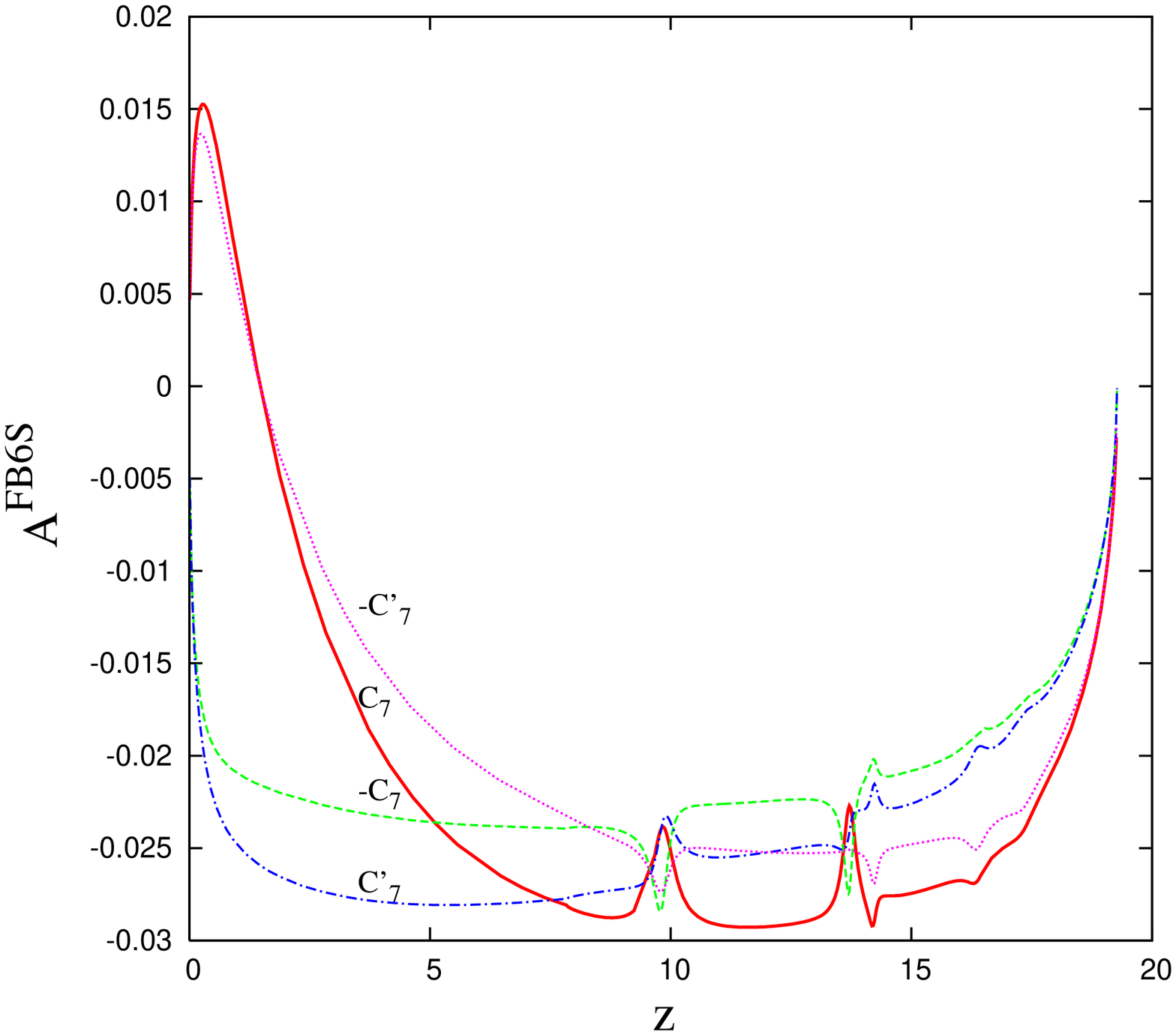}}
\end{minipage}
    \hspace*{10mm}
\begin{minipage}[c]{0.4\textwidth}
\end{minipage}
\end{center}
\caption{The asymmetries defined in
Eqs. (\ref{FBs2})-(\ref{FBs6}), where the solid (red) line shows the SM
case with scalar resonance, the dashed (green) shows
the $-C_7$ case, the dash-dotted (blue) line is the pure $C_7^\prime
=|C_7|$ case and the dotted (purple) line is for
$-C_7^\prime $ case.  Here we did not assume any new CP phase.}
\end{figure}
In Fig. 7, we show the dependence of new CP
phase for $FB_2^s$, $FB_3^s$ and $FB_4^s$,  $A^{FB_i^s} $ and $A_{CP}^{FB_i^s}$
as functions of $z$, where the new phases of $C_9$
(for $FB_2^s$) and $C_{10}$ (for $FB_{(3,4)}^s $)
are taken as $0, \pi/8, \pi/4 $ and $\pi/2 $.
The direct CP asymmetries at low $z$ region seem to be enhanced by  strong phase
differences induced by interferences with the scalar
resonance. Unfortunately, $A_{CP}^{FB_2^s}$ will be quite small because it is
proportional only to $C_9^{eff *} C_7$ term. However,
$A_{CP}^{FB_3^s}$ and $A_{CP}^{FB_4^s}$ are very interesting at low $z$
region because we see the enhancement effects through the interference. And the
contributions from the nonstandard interactions $C_i^\prime $ can be
enhanced in newly defined CP asymmetries.

\begin{figure}[htbp]
\begin{center}
\begin{minipage}[c]{0.4\textwidth}
{\includegraphics[scale=0.35,angle=-90]{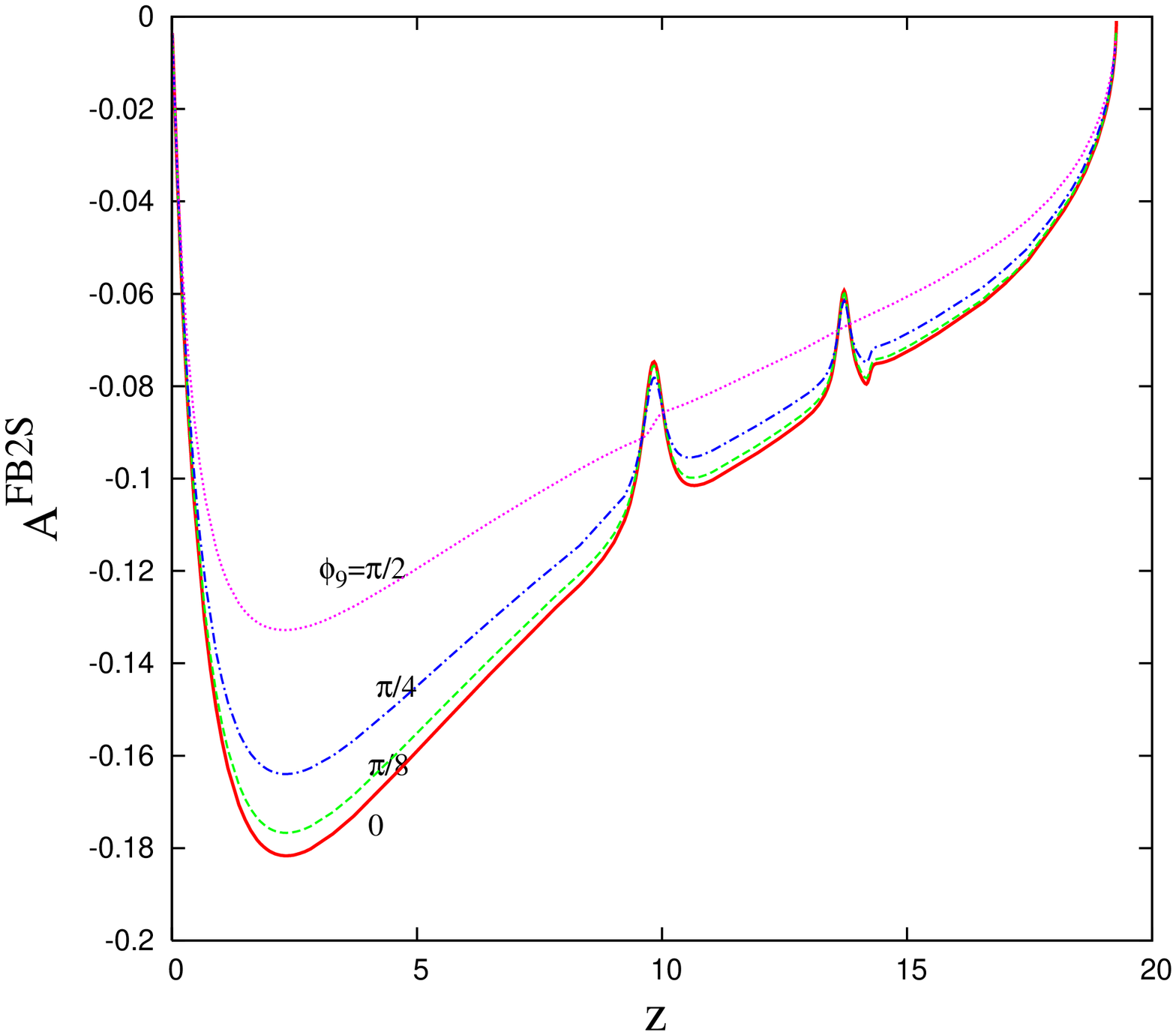}}
\end{minipage}
    \hspace*{10mm}
\begin{minipage}[c]{0.4\textwidth}
{\includegraphics[scale=0.35,angle=-90]{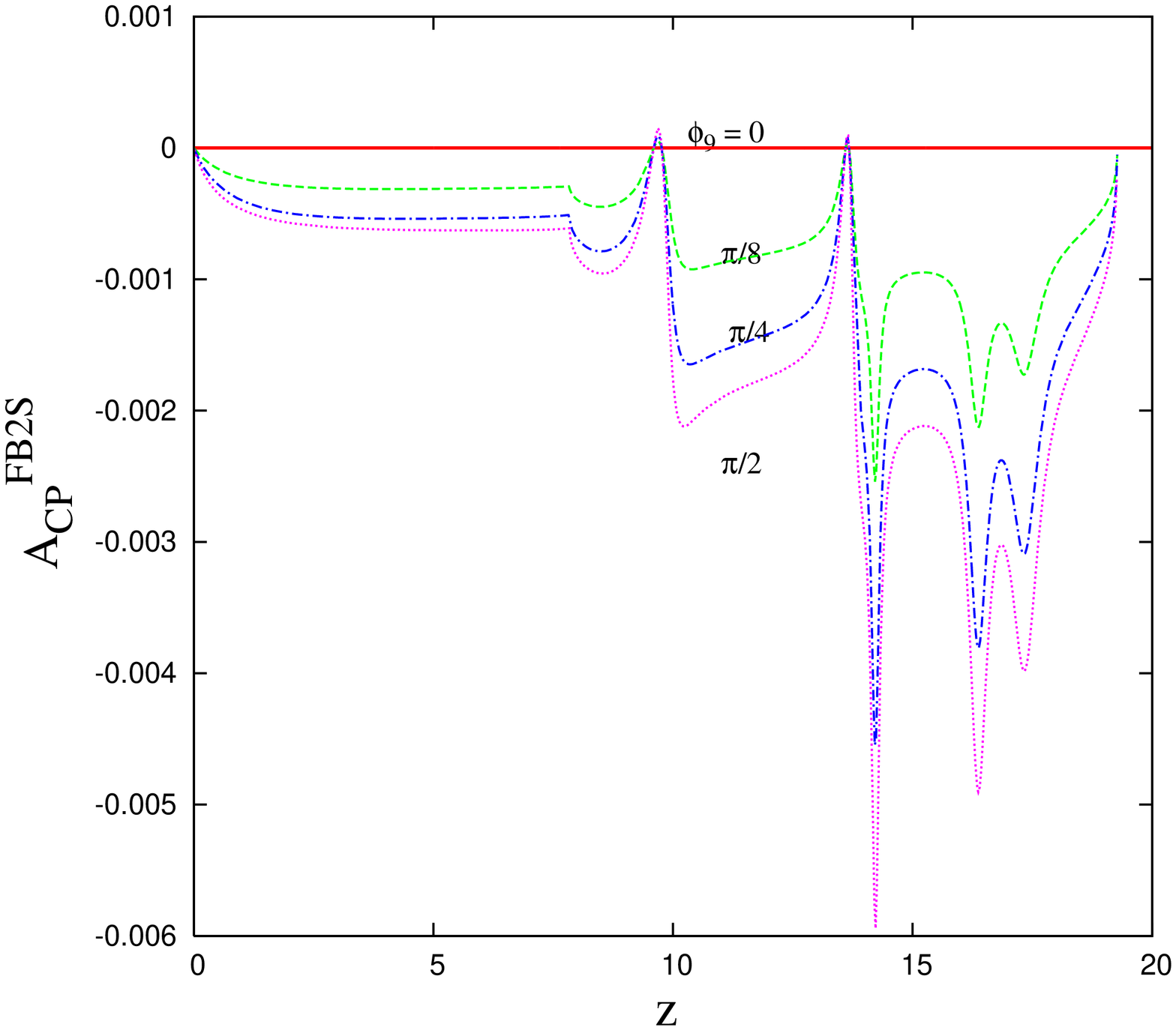}}
\end{minipage}\\[5mm]
\begin{minipage}[c]{0.4\textwidth}
{\includegraphics[scale=0.35,angle=-90]{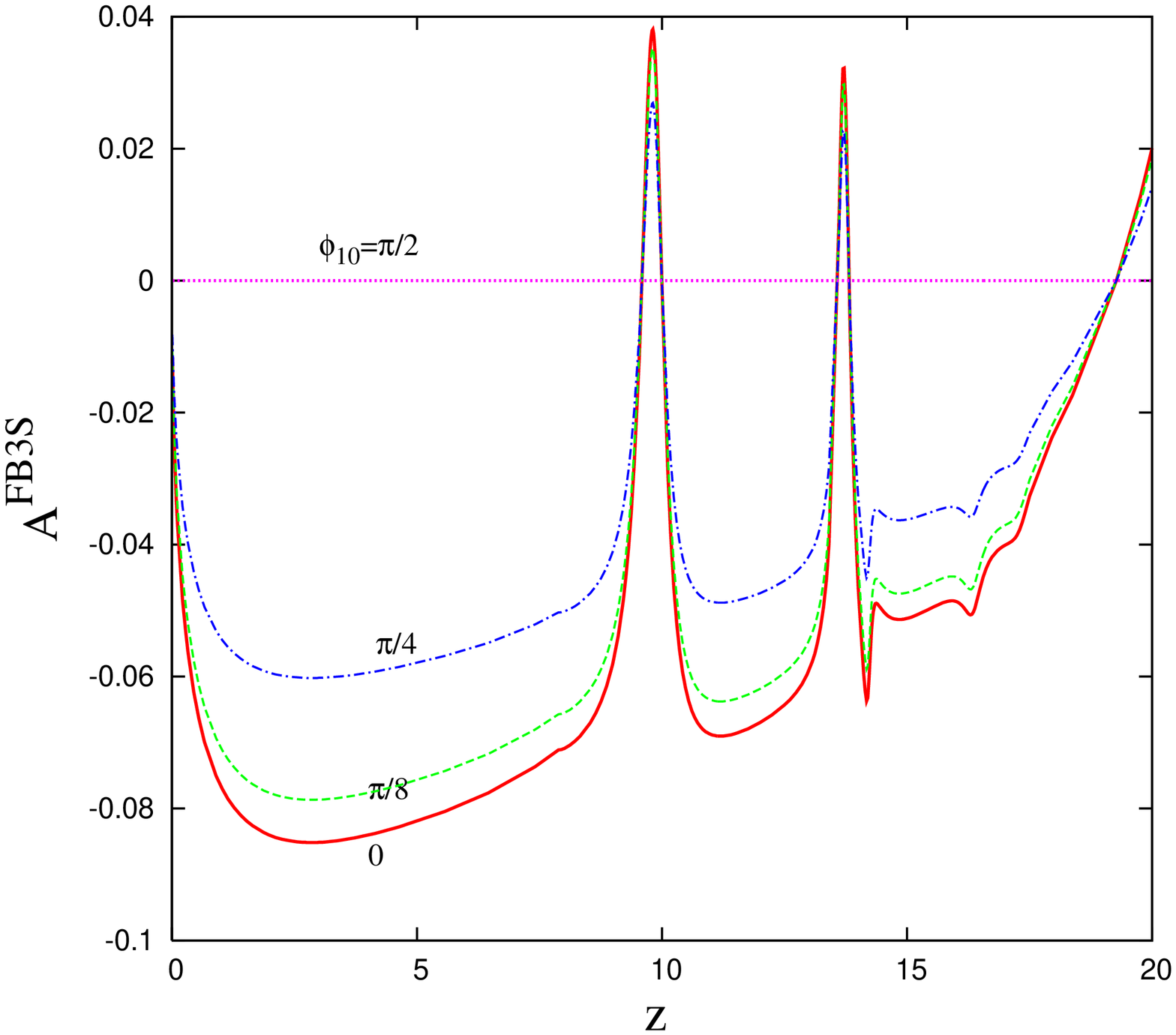}}
\end{minipage}
    \hspace*{10mm}
\begin{minipage}[c]{0.4\textwidth}
{\includegraphics[scale=0.35,angle=-90]{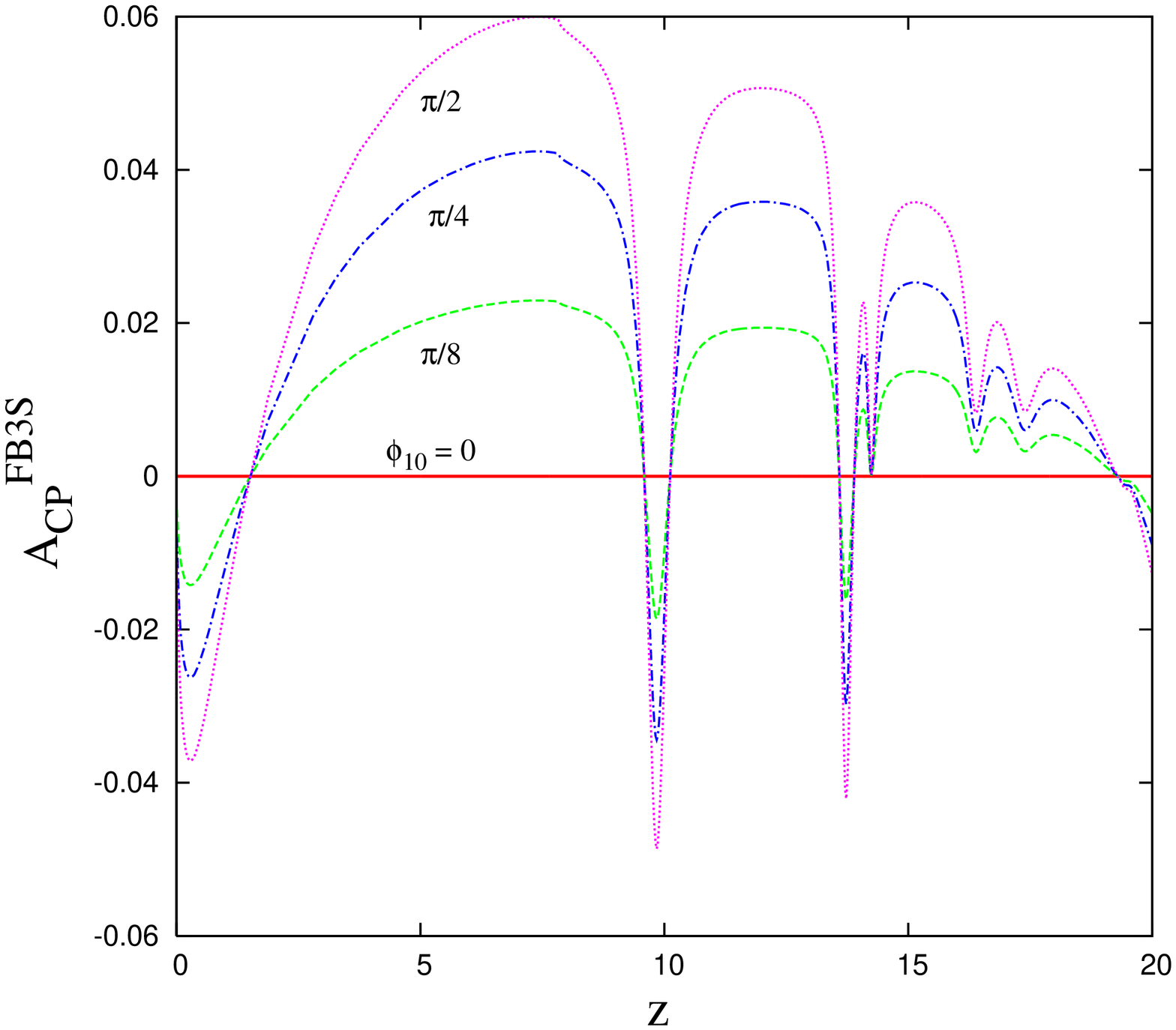}}
\end{minipage}\\[5mm]
\begin{minipage}[c]{0.4\textwidth}
{\includegraphics[scale=0.35,angle=-90]{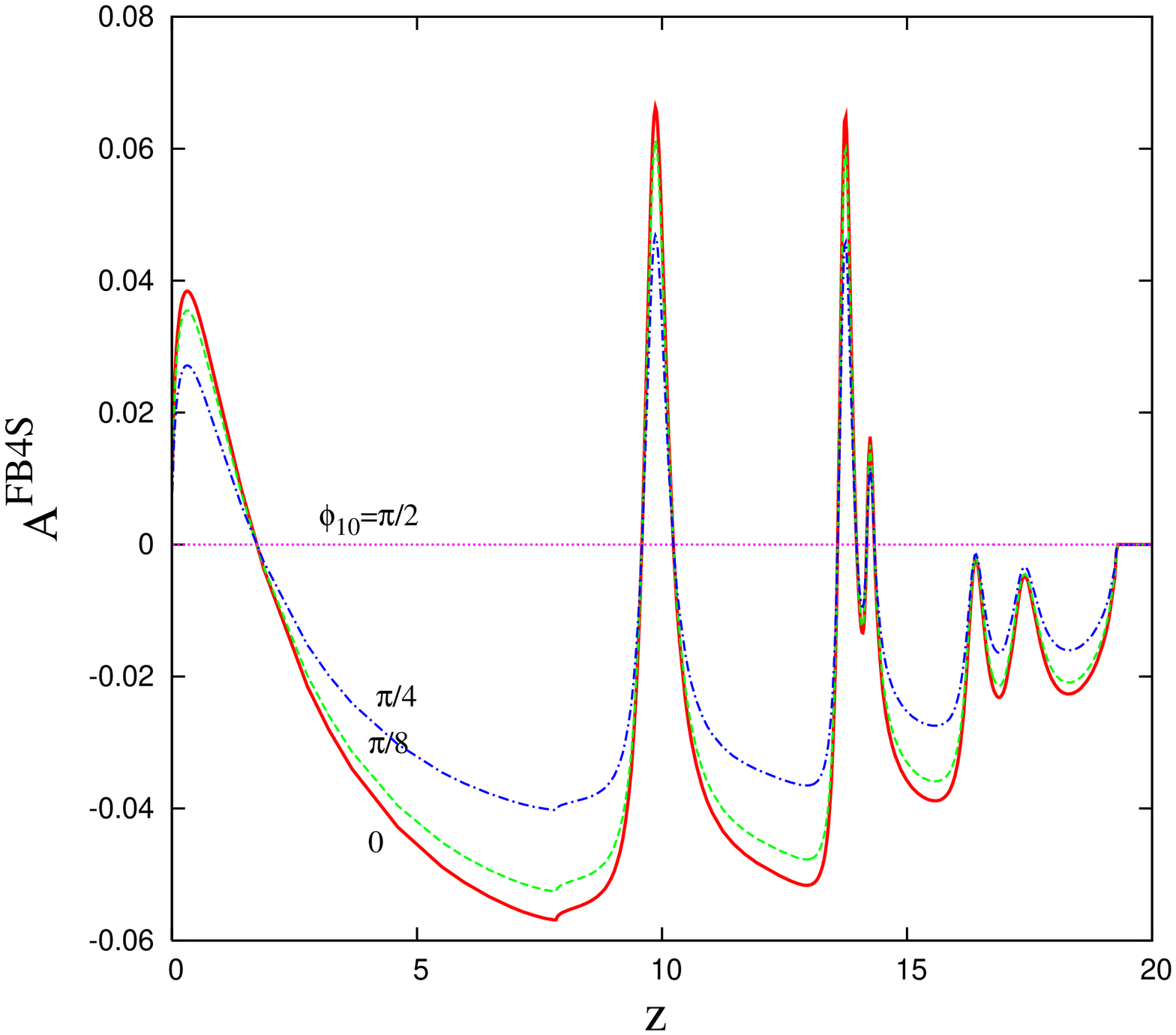}}
\end{minipage}
    \hspace*{10mm}
\begin{minipage}[c]{0.4\textwidth}
{\includegraphics[scale=0.35,angle=-90]{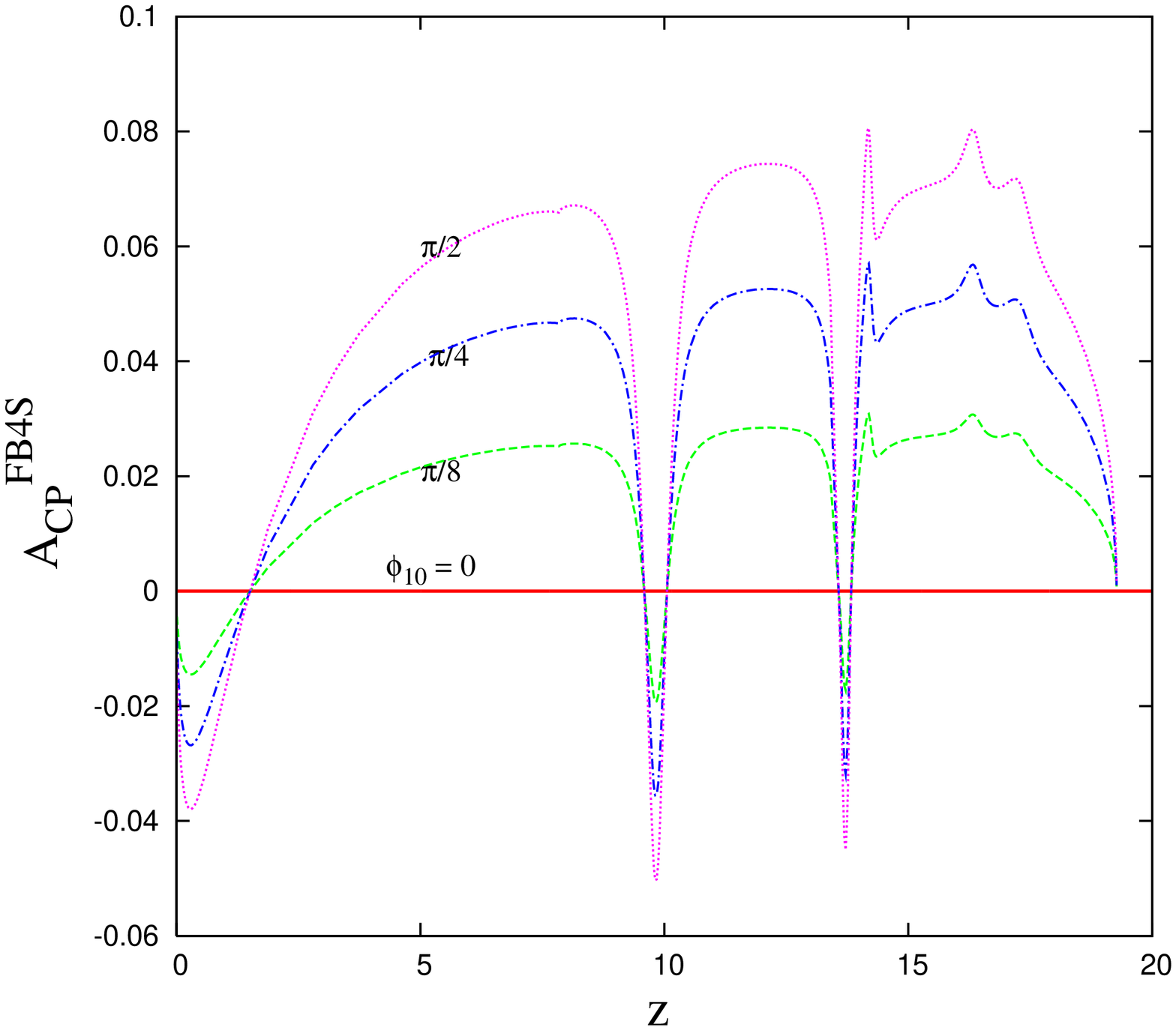}}
\end{minipage}
\end{center}
\caption{$A^{FB_i^s} $ and $A_{CP}^{FB_i^s}$
are plotted as functions of $z$, where new phases of $C_9$
($A^{FB_2^s}$) and $C_{10}$ ($A^{FB_3^s}$ and $A^{FB_4^s}$)
are taken as $0, \pi/8, \pi/4 $and $\pi/2 $.}
\end{figure}

Nonzero values for newly defined FB asymmetries will indicate strong
evidences for the existence of the scalar resonance in addition to vector $K^*$ meson.
Indeed, future super-$B$ factories and LHC-b experiment can
measure and count the events for some regions of phase space after separating several
bins. Surely we can detect such contributions from these measurements.
If we find these contributions quite large,
the interferences may have
an important role as a source of strong phase difference, which
is one of the conditions to enhance CP asymmetries.

\section{Summary}

Based on the most general 4-fermi interaction, which includes all types
of possible interactions with new CP phases, we have investigated the general 4-body
decay process, $B\rightarrow K\pi l^+ l^- $, through the angular decomposition method.
As is well known, this 4-body decay process can be described in general by using
3 angles, so that we can extract many useful information from
the angular decomposition analysis. Similar to $B \to K^* l^+ l^-$, we can probe the region of
zero point for the leptonic  FB asymmetry, $A_{FB}=0$, as well. However, here in this general
4-body analysis we can obtain much more information to extract the sources of new physics.
We can define several  CP averaged FB asymmetries, direct CP asymmetries as well as
time dependent CP asymmetries as functions of the 3 angles in terms of
the general 4-fermi interaction parameters.
We found that some of them are very sensitive to
strong or EW phases, and
some of them are from interference contributions between CP odd
and CP even modes so that the CP asymmetry can be enhanced.

Note that for the decays of $B \to M [\to K \pi]~ l^+ l^-$,
if we do not constrain the invariant mass of
$K$-$\pi$ system, there exist several intermediate mesons contributing to
$B \to K \pi l^+ l^-$ decays. Therefore, through the interference we may induce large strong phases,
which result possibly large CP violations if there exist any new physics
CP phases beyond the CKM phase. We considered the case with the scalar resonance decay
$B \to K_0^*(800) [\to K \pi]~ l^+ l^-$ in addition to the vector resonance decay
$B \to K^*(892) [\to K \pi]~ l^+ l^-$. Again we can define new type of
several FB asymmetries and direct CP asymmetries resulted from
the interference of vector and scalar intermediate mesons.
We investigated the interference effects as a source
of strong phase difference, the same as imaginary part of
$C_9^{eff}$ within the SM, to obtain a few hints of new physics effects.
By considering these asymmetries systematically,
we can obtain several hints for
new CP phases in EW penguin decays, and find that the angular decomposition analysis
for the general 4-body decay process can be very
useful tool to understand new physics, which may be hiding in EW penguin.
If the interference effect is fortunately quite large,
we can use it as an enhancement of CP asymmetries to find new CP
phases very clearly.

Future super-$B$ factories \cite{superB} and LHC-b
may be able to find out unknown resonance states and
investigate the dependence of new physics in detail.
At very low region of dilepton invariant mass \cite{GrossPirjol}
by using photon
conversion technique  \cite{Ntech}, the new
contribution from right-hand current and the CP phase of
$C_7^{\{\prime \}}$ type interaction may be measured. Then, we have to
consider more carefully on measuring  the angular distributions of
$B\rightarrow K\pi l^+ l^- $ and the related CP asymmetries to find out
information of not only
new CP phases of $C_9 $ and $C_{10}$ type interactions but also
the most general 4-fermi interaction type new physics.
Hence, we expect
our analysis will be very useful to find new physics hiding beyond
$C_9$ and $C_{10}$ with new CP violating phases.

\section*{Acknowledgments}

The work of C.S.K. was supported
in part by  CHEP-SRC Program,
in part by the Korea Research Foundation Grant funded by the Korean Government (MOEHRD) No. KRF-2005-070-C00030.
The work of T.Y. was supported by 21st Century COE Program of
Nagoya University.
\\


\end{document}